%% file: chess-u_accelerator_design.tex
\documentclass[aps,prstab,twocolumn,groupedaddress,
nofootinbib,longbibliography,floatfix]{revtex4-1}
\usepackage{graphicx}
\usepackage{dcolumn}
\usepackage{etoolbox, ifpdf}
\usepackage{color}
\usepackage{soul}
\usepackage{amsmath}
\usepackage{subcaption}
\usepackage[export]{adjustbox}
\usepackage{hyperref}
\hypersetup{
    pdfstartview={FitH},
    colorlinks=false,
    pdfborder={0 0 0},
    breaklinks=true
}
\usepackage{breakurl}

\begin{document}

\title{Accelerator Design for the CHESS-U Upgrade}
\author{J. Shanks}
\email{shanks@cornell.edu}
\author{J. Barley}
\author{S. Barrett}
\author{M. Billing}
\author{G. Codner}
\author{Y. Li}
\author{X. Liu}
\author{A. Lyndaker}
\author{D. Rice}
\author{N. Rider}
\author{D.L. Rubin}
\author{A. Temnykh}
\author{S.T. Wang}
\affiliation{CLASSE, Cornell University, Ithaca, New York 14853,
USA}

\date{\today}

\begin{abstract}
During the summer and fall of 2018 the Cornell High Energy
Synchrotron Source (CHESS) is undergoing an upgrade to increase
high-energy flux for x-ray users. The upgrade requires replacing
one-sixth of the Cornell Electron Storage Ring (CESR), inverting the
polarity of half of the CHESS beam lines, and switching to
single-beam on-axis operation. The new sextant is comprised of six
double-bend achromats (DBAs) with combined-function
dipole-quadrupoles. Although the DBA design is widely utilized and
well understood, the constraints for the CESR modifications make the
CHESS-U lattice unique. This paper describes the design objectives,
constraints, and implementation for the CESR accelerator upgrade for
CHESS-U.
\end{abstract}

\maketitle

\section{Introduction}

\begin{figure*}[htb] 
   \centering
   \includegraphics[width=0.45\textwidth]{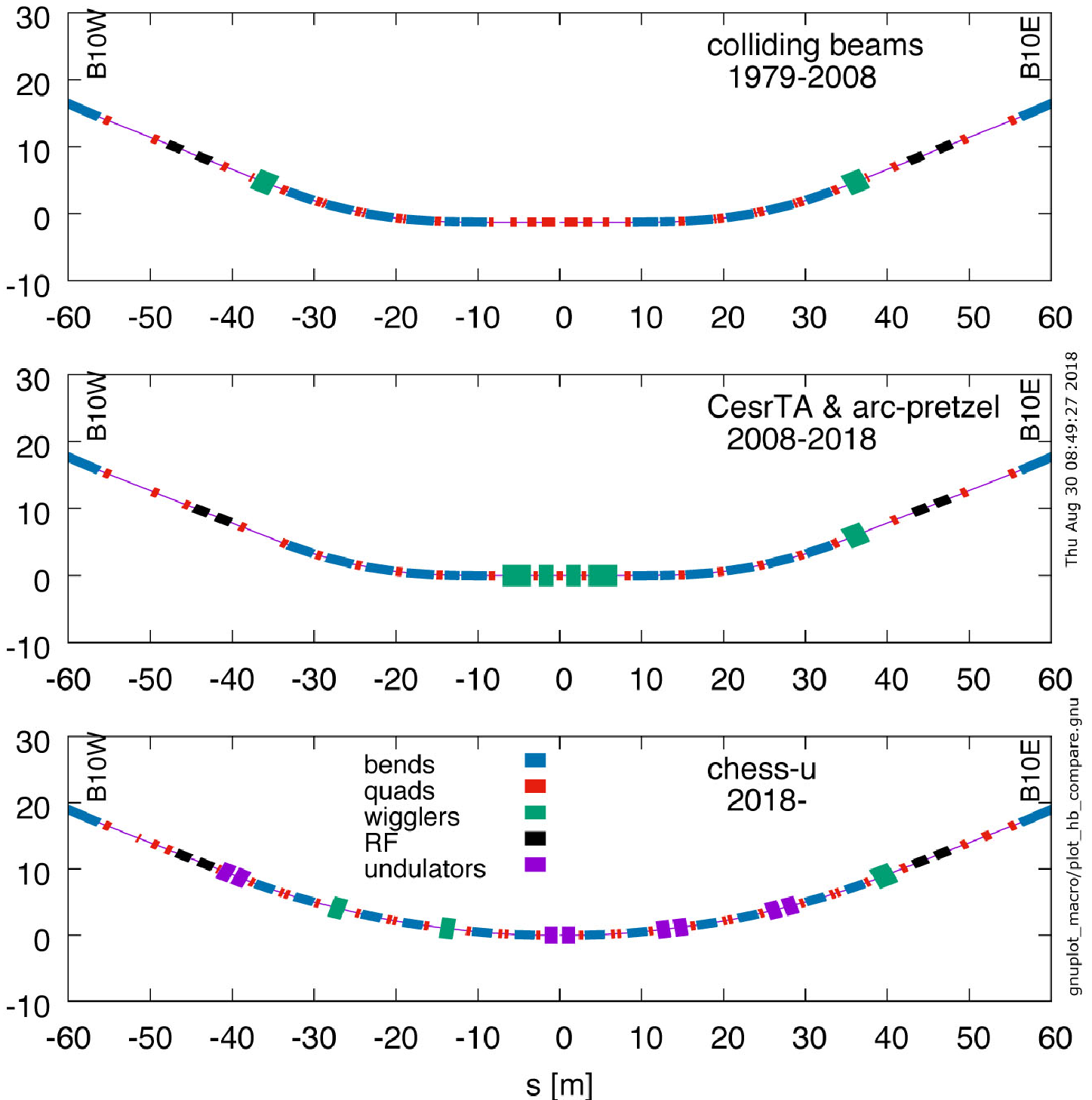}
   \includegraphics[width=0.45\textwidth]{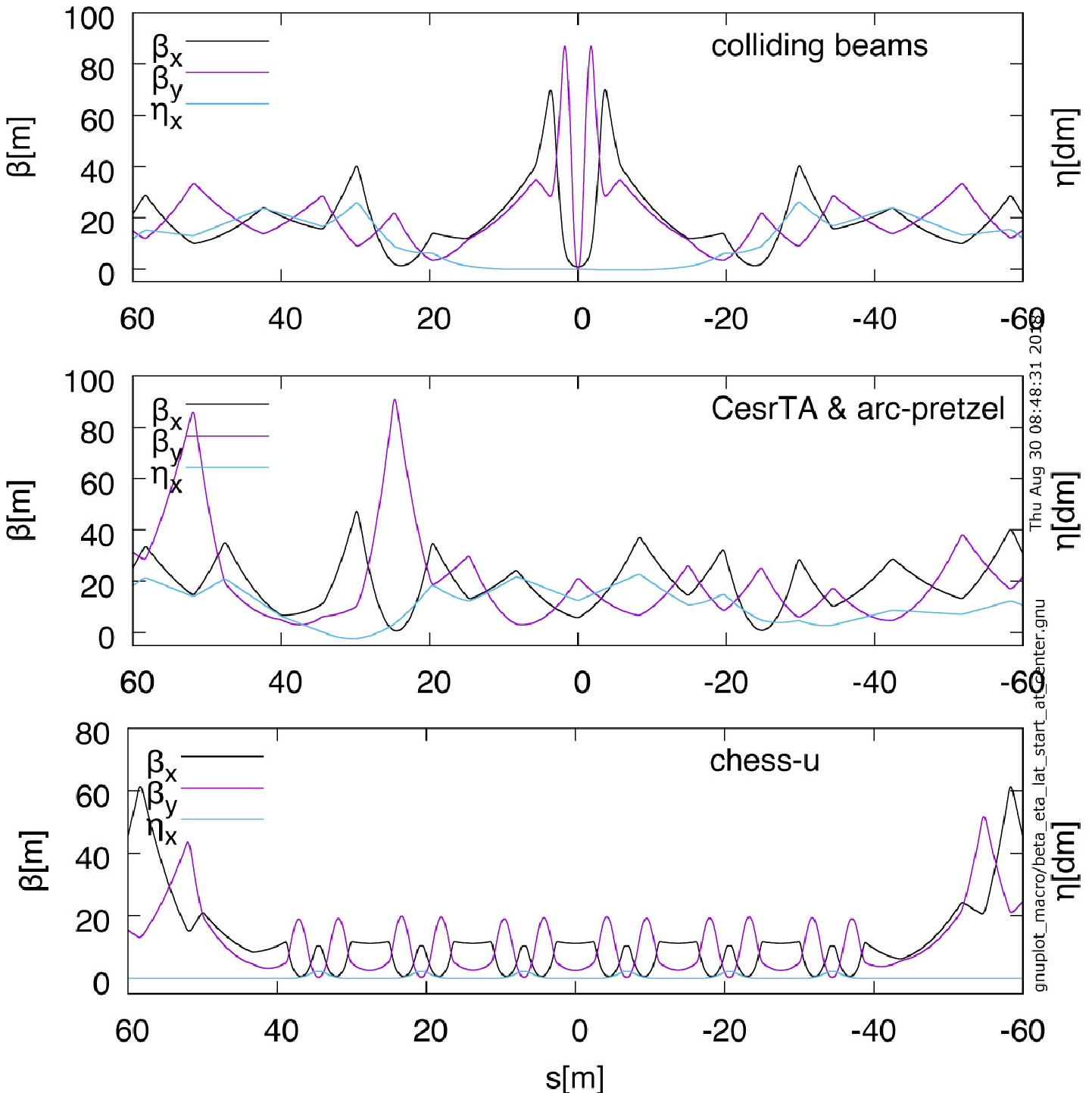}
   \caption{\label{fig:layout} At left is the magnetic layout of the L0 arc
   of the storage ring; top) Colliding beams, middle)
   CesrTA and CHESS arc-pretzel, bottom) CHESS-U. At right optics in the
L3 straight for colliding beams, arc-pretzel and CHESS-U.}
\end{figure*}

The Cornell Electron/positron Storage Ring (CESR) was designed as an
electron-positron collider, with two diametrically opposed
interaction regions. The layout of the storage ring is constrained
by the very nearly circular geometry of the tunnel that it shares
with the existing synchrotron. The CLEO detector required a long
straight section to accommodate strong IR quads for a mini-beta
insert and anti-solenoids to compensate the 1.5~T CLEO solenoid, in
addition to the detector itself. Hard-bend dipoles were required to
bend the trajectory into the long straight, with gentler bends
immediately adjacent to the straight to mitigate synchrotron
radiation background in CLEO. The collider layout and characteristic
optics are shown in Fig. ~\ref{fig:layout}.

The Cornell High Energy Synchrotron Source (CHESS), founded
contemporaneously with CESR, was initially comprised of three x-ray
extraction lines off the hard-bend dipoles from the
counter-clockwise electron beam, expanding in 1988-89 with the
addition of four beam lines (including one permanent-magnet wiggler)
off the clockwise positron beam in CHESS East, and again in 1999
with the construction of G-line, adding three end stations fed by an
insertion device from the positron beam.

Following the conclusion of the CLEOc HEP program in 2008, CESR was
reconfigured for the CESR Test Accelerator program (CesrTA)
\cite{JINST10:P07012} damping ring R\&D program. The CLEO particle
detector drift chamber was removed and the final focus replaced with
a conventional FODO arrangement. The long straight was outfitted
with six superconducting damping wigglers, relocated from two
straights in the arcs.

The CesrTA layout also enabled the operation of CHESS in the
arc-pretzel configuration with counter-rotating beams of electrons
and positrons \cite{Wang:IPAC2016-WEPOW053}. Layout and optics are
shown in the middle plots in Fig. ~\ref{fig:layout}.

CHESS was designed to run parasitically with HEP operations, with
beam lines fed by both electron and positron beams. Running with two
electrostatically-separated beams greatly complicates many facets of
operation, including restricting bunch patterns and maximum
per-bunch current, increasing the tune plane footprint, and
impairing the dynamic and momentum apertures.

With the conclusion of HEP operations, CESR is being reconfigured as
a dedicated high-energy, high-performance single-beam synchrotron
light source, CHESS-U. The conversion to operation with a single
clockwise beam requires rebuilding the beam lines fed by the
counter-clockwise beam. The new layout eliminates the long IR
straight and hard bends in favor of evenly distributed achromats and
insertion straights, as shown in Fig. \ref{fig:layout}.

\section{Lattice Design}
\input{s2_lattice_design.tex}


\section{Lattice Characterization}
\input{s3_lattice_characterization.tex}

\section{Magnet Design}
\input{s4_magnet_design.tex}


\section{Vacuum Design}
\input{s5_vacuum_design.tex}


\section{Instrumentation}

\input{s6_instrumentation.tex}

\section{Bunch Patterns}
\input{s7_bunch_patterns.tex}

\section{Installation and Commissioning}
\input{s8_commissioning.tex}

\section{Summary}
    CHESS-U will deliver high flux photon beams at high energy, enabling
    a diverse set of new experiments \cite{CHESSU:ScienceCase}.
    The upgrade will improve the performance of the
    CESR storage ring for x-ray production by almost an order of
    magnitude with CCUs and a factor of three for the 24-pole
    ``F-line'' wiggler. In turn, the transition to single-beam will
    drastically simplify operations. Bunch patterns compatible with
    timing mode are under consideration, pending evaluation during
    commissioning and machine studies.

\section{Acknowledgements}

    The authors wish to thank everyone involved in the CHESS-U
    upgrade project, and those who have contributed material to
    this paper.

    Funding for the CHESS-U upgrade provided by New York State
    Capital Grant \#AA737 / CFA \#53676.


\appendix
\input{appendices.tex}

\clearpage
\bibliography{chessu}

\end{document}

%% file: s2_lattice_design.tex
\subsection{Design Objectives and Constraints}

The accelerator requirements for the CHESS-U upgrade are: 1)
Minimize the natural emittance for a single on-axis beam at 6.0~GeV;
2) Introduce straights for independent insertion devices for most
end stations; 3) Preserve compatibility with existing injector
system (linac and synchrotron booster ring); and 4) Increase
high-energy x-ray flux for all end stations.

\begin{figure*}[t]
\centering
    \begin{subfigure}[t]{0.85\textwidth}
        \includegraphics[width=\textwidth]{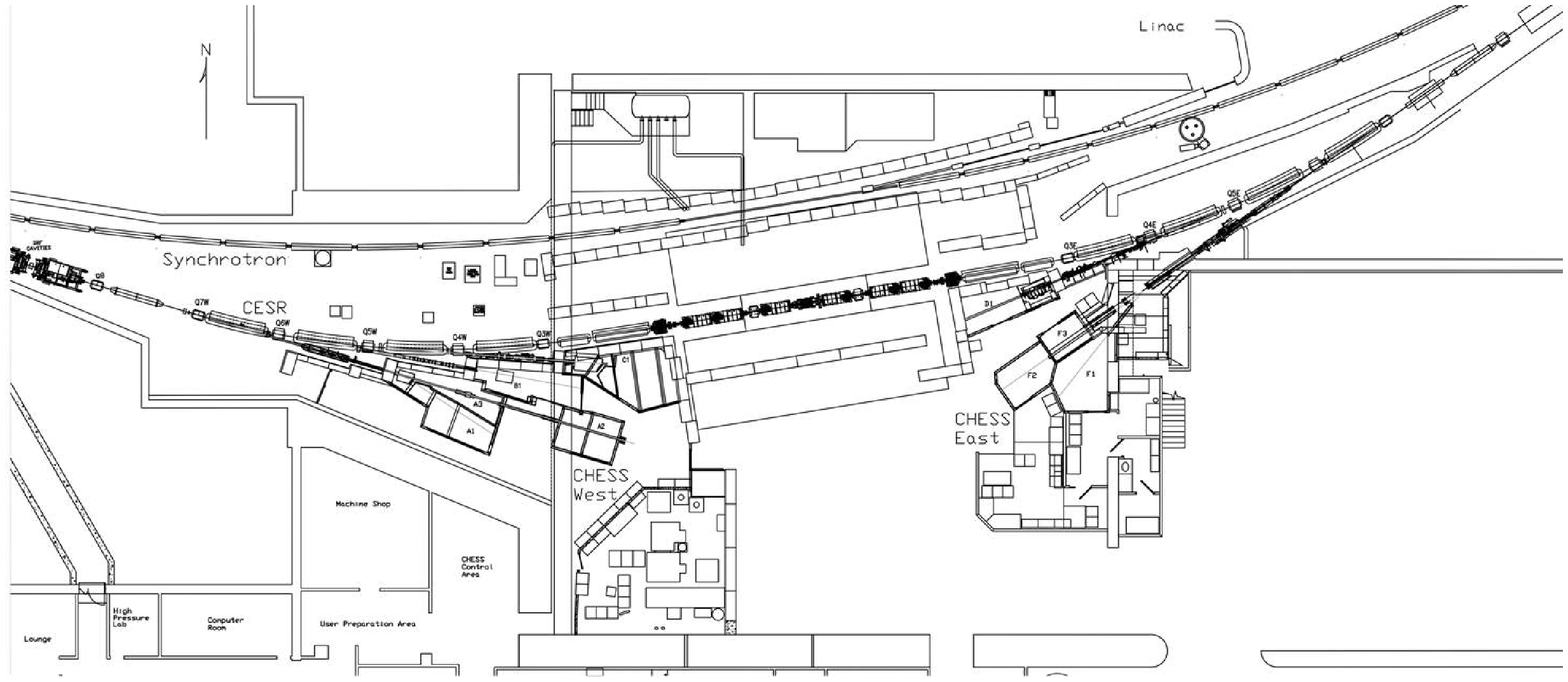}
        \caption{Layout before CHESS-U.}
    \end{subfigure}
    \begin{subfigure}[t]{0.85\textwidth}
        \includegraphics[width=\textwidth]{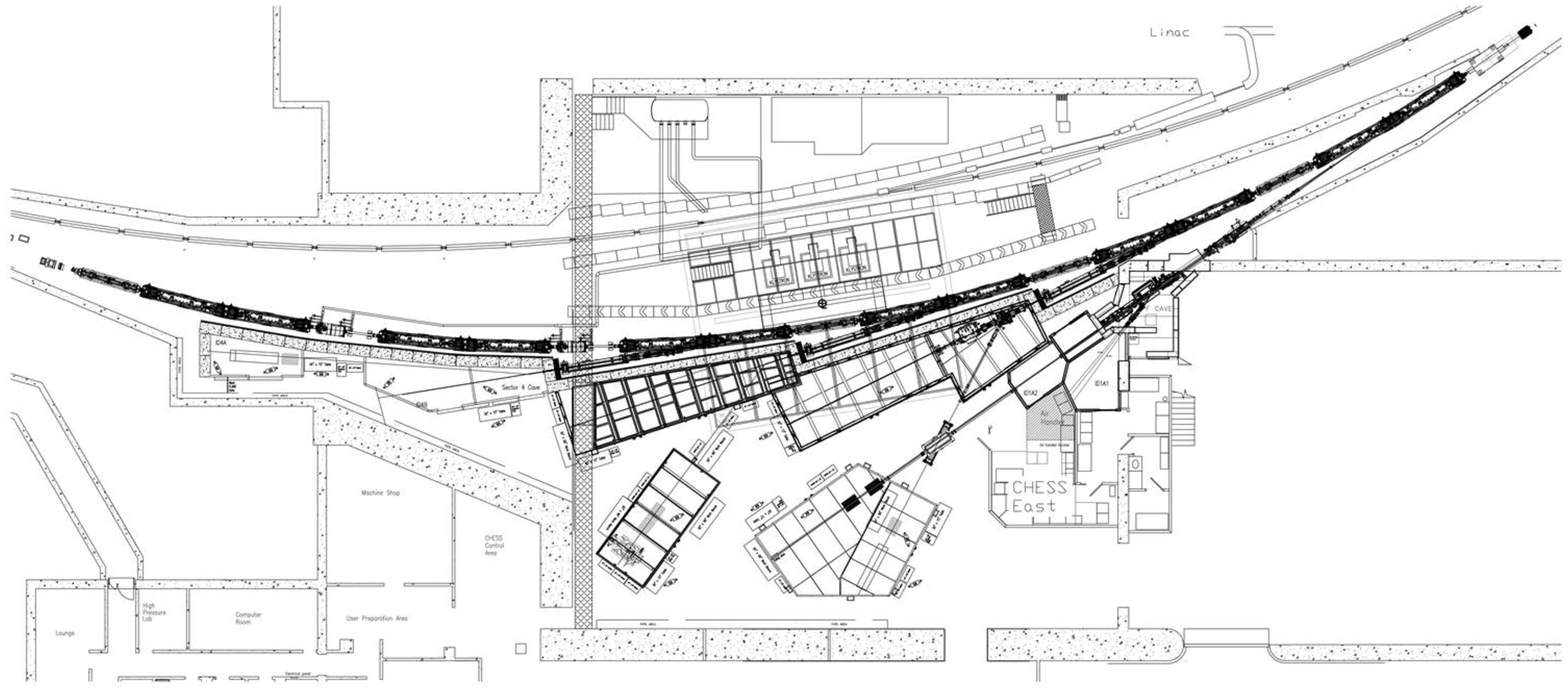}
        \caption{CHESS-U layout.}
    \end{subfigure}
    \caption{CHESS layout, before and after CHESS-U upgrade.}
    \label{fig:chess_floor}
\end{figure*}

Most of the CHESS beam lines are presently located in the L0 main
experimental hall, fed by both electron and positron beams, see Fig.
\ref{fig:chess_floor}. Eight end stations are located in L0, four
from each beam; a further three end stations were added in 1999 as
an excavation out of the west CESR tunnel, fed by the
clockwise-oriented positron beam. For single-beam operation, a
clockwise orientation was chosen in order to preserve the existing
six end stations from the F-line and G-line facilities.

\begin{figure}
\centering
    \includegraphics[width=0.45\textwidth]{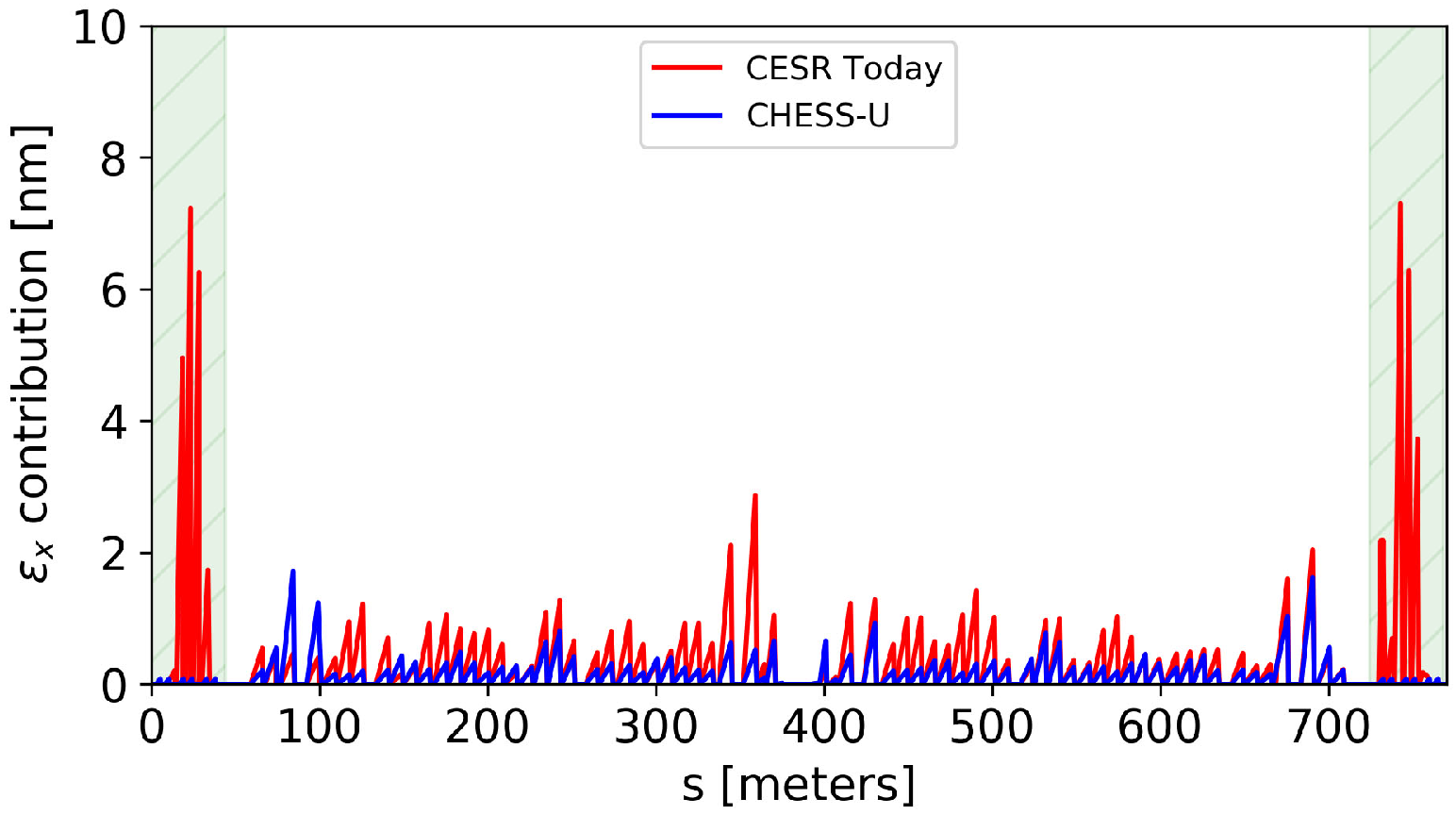}
    \caption{Element-by-element contribution to the natural emittance,
    via the $I_5$ radiation integral. The green hatching indicates
    the region to be upgraded for CHESS-U.
    $s=0$ corresponds to the center of the former CLEO interaction region,
    which is also the center of the CHESS user region of the storage
    ring. The largest contributions to $I_5$ in the existing CESR layout correspond to the hard
    bend dipoles which bend into the long CLEO IR straight.}
    \label{fig:i5_per_ele}
\end{figure}

The existing storage ring layout in L0 includes one long straight
section for the former CLEO detector, which is not optimal for
low-emittance x-ray light source operation. The present optics do
not have enough degrees of freedom to constrain the dispersion in
the hard bend dipoles which bring CESR into the long straight, and
as a result, the hard bends dominate the natural emittance (see Fig.
\ref{fig:i5_per_ele}). Additionally, the long straight section
cannot accommodate multiple source points for x-ray beam lines,
which require angular separation between sources.

The L0 experimental hall was also home to the three-story CLEO
particle detector, which ran for HEP operation from 1979-2008. For
single-beam CHESS operation, several x-ray extraction lines would
pass through the CLEO iron; to make room for new source points, the
CLEO detector was disassembled and removed in 2016.

\subsection{Linear Optics}

In order to maximize the number of end stations, a layout was chosen
with six 13.8~m double-bend achromats with 3.5~m insertion device
(ID) straights. It is worth noting the cell length is approximately
half that of many third-generation storage ring light sources. DBAs
were chosen over a multi-bend achromat design for a number of
reasons: 1) The space available is extremely limited; 2) The dynamic
aperture requirements would be incompatible with the existing
accumulation injection scheme; and 3) the cost and complexity would
be beyond the scope of this project.

The new DBAs utilize 2.35~m defocusing combined-function sector
dipole-quadrupoles (DQs). Though the bend radius of the new CHESS-U
dipoles is nearly identical to the old hard bends (31.4~m vs.
31.7~m), the dispersion is well suppressed in the new achromats,
leading to a factor of four reduction in the global emittance while
only replacing one-sixth of the storage ring.

\begin{figure}
\centering
    \includegraphics[width=0.45\textwidth]{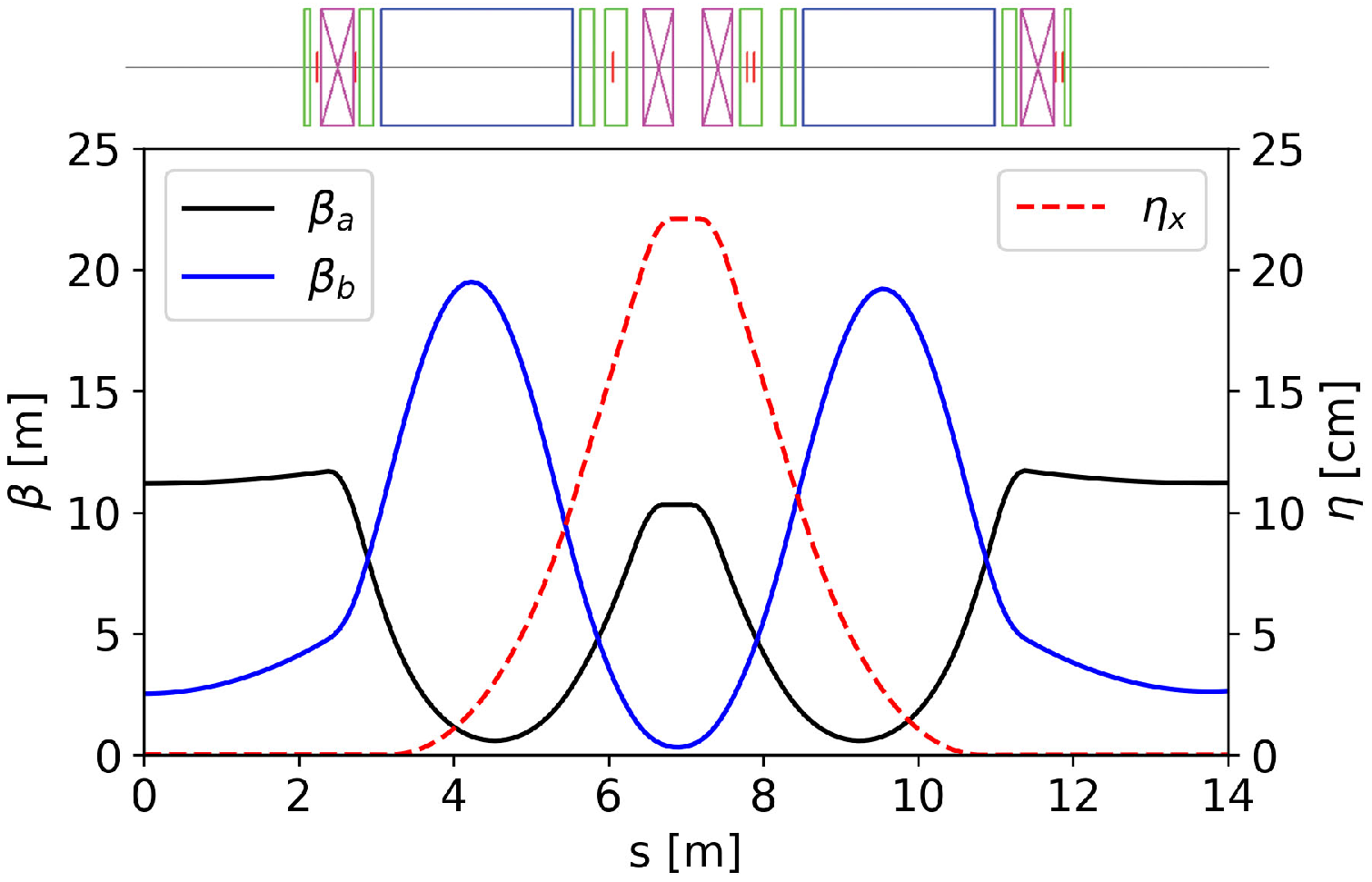}
    \caption{One DBA cell for the CHESS-U upgrade. Blue boxes are
    DQs, purple are quadrupoles, green
    are steerings. Red ticks are beam position monitors.}
    \label{fig:dba}
\end{figure}

\begin{figure}
\centering
    \includegraphics[width=0.45\textwidth]{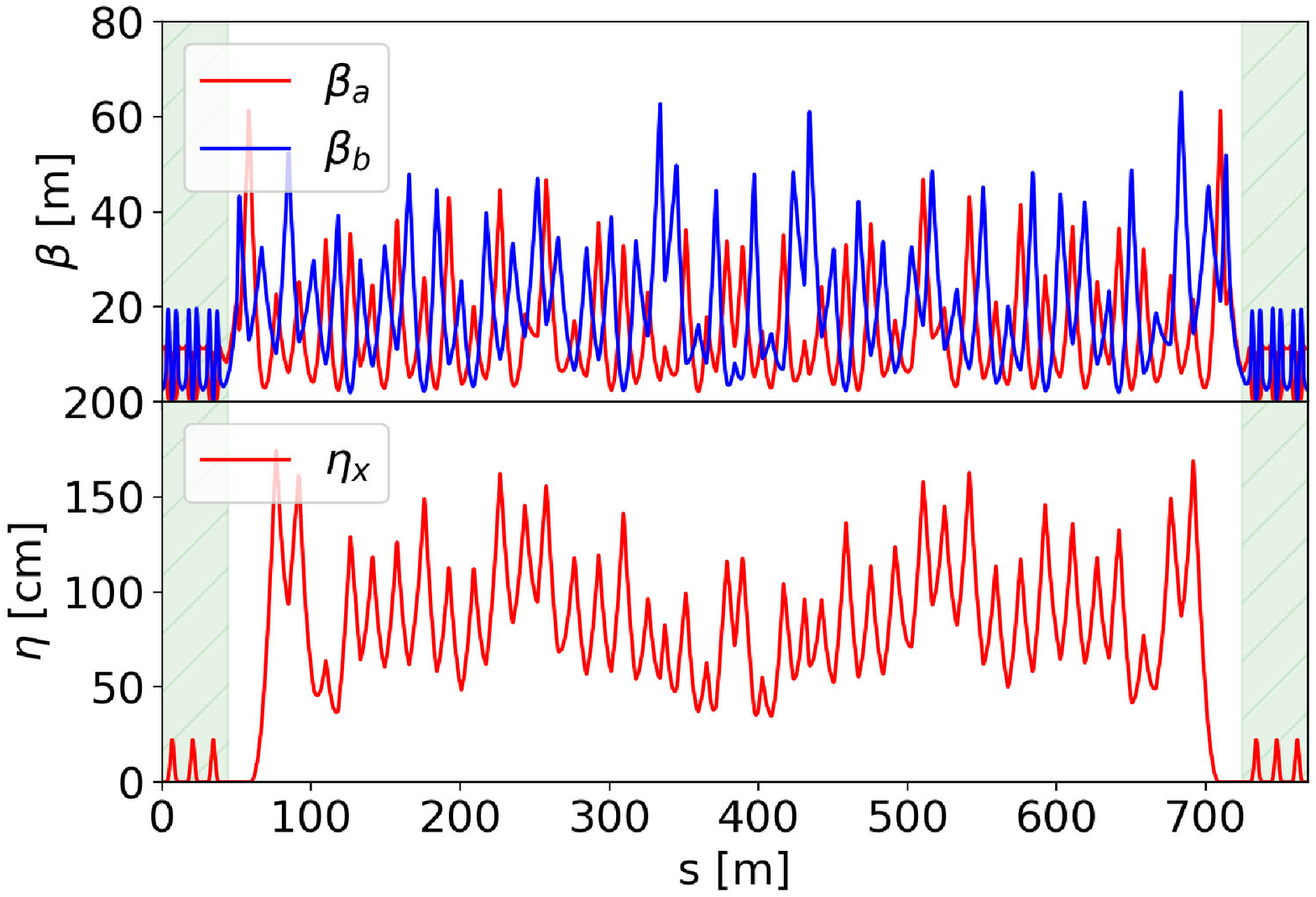}
    \caption{Full ring optics for CHESS-U. $s=0$ corresponds to the
    center of the former CLEO interaction region, which becomes the
    center of the Sector 4 insertion device straight. The green hatching
    indicates the region being upgraded for CHESS-U.}
    \label{fig:full_ring}
\end{figure}

A single DBA shown in Fig. \ref{fig:dba}. The natural emittance for
a single CHESS-U cell is 2.56~nm$\cdot$rad at 6.0~GeV. Matched into
the remainder of the CESR, the full ring optics are shown in Fig.
\ref{fig:full_ring}. Linear and nonlinear optics were optimized
using Tao \cite{PAC05:FPAT085}, which is built on the Bmad
accelerator library \cite{NIMA558:356to359}.

The six new achromats span the 100-meter region between the two
superconducting rf straights, and begins and ends with straights for
insertion devices. Five of the seven straights will be used for
CHESS IDs; there is not presently space for end stations for the
remaining two straights. The remaining two straights will have one
CESRc superconducting damping wiggler each \cite{pac01:3648M}, to be
used only in low-energy machine studies. The new CHESS-U lattice
will enforce zero dispersion through the rf cavities, a condition
which is not met in present two-beam CHESS optics. The new achromats
are matched into the existing FODO structure with four additional
quadrupoles in each of the rf straights.

The storage ring energy will be increased to 6.0~GeV. Aside from
CLEOc HEP operation in 2003-2008 and occasional CesrTA machine
studies from 2008-2016, when the energy was lowered to
2.085~GeV/beam, CESR has operated at 5.3~GeV/beam. CHESS operation
has been exclusively at 5.3~GeV. CESR was originally designed for an
energy range of up to 8.0~GeV, though the highest operating energy
to-date is 5.6~GeV. Machine studies conducted in 2017 in the present
CESR configuration demonstrated injection and accumulation at
6.0~GeV. Every quadrupole and sextupole in CESR is independently
powered, allowing significant flexibility in operating conditions.
Therefore the remaining 5/6 of the storage ring requires no
modification.

CHESS-U lattice parameters are summarized in
Table~\ref{tab:lattice_params}.

\begin{table}
 \begin{centering}
     \begin{tabular}{cccc}
         \toprule
         \textbf{Parameter}    & \textbf{CHESS}    & \textbf{CHESS-U}    & \textbf{Units}\\
         \colrule
         Circumference         & 768.438           & 768.438             & m \\
         Energy                &  5.289            & 6.0                 & GeV  \\
         Species               & e$^+$ and e$^-$   & e$^+$               & -- \\
         Current               & 120/120           & 200                 & mA \\
         $\epsilon_x$          & 98                & 29.6                & nm$\cdot$rad  \\
         Emittance coupling    & 1\%               & 1\% & --\\
         $\beta_{x,y}$ at IDs  & 7.9, 3.1          & 11.2, 2.6           & m      \\
         $\eta_x$ at IDs       & 0.42              & 0                   & m \\
         IDs                   & 3                 & 9                   & -- \\
         End Stations          & 11                & 12                  & -- \\
         $Q_{x,y}$             & 11.28, 8.78       & 16.55, 12.63        & -- \\
         $Q^{\,\prime}_{x,y}$       & -15.95, -14.20    & -25.64, -26.76      & -- \\
         $\alpha_p$            & $9.2\times 10^{-3}$ & $5.7\times 10^{-3}$ & -- \\
         Bunch Length          & 16                & 14                  & mm \\
         Per-bunch Current     & 7                 & 2.2                 & mA \\
         Touschek Lifetime     & $>$24               & 40                   & hrs\\
         RF Voltage            &     5.2           &    6       &       MV\\
         \botrule
     \end{tabular}
     \caption {Lattice parameters for CHESS today and CHESS-U upgrade.}
     \label{tab:lattice_params}
 \end{centering}
 \end{table}

\subsection{Nonlinear Optics}

Due to the lack of periodicity in CESR (the nearest symmetry in the
ring is a mirror symmetry about the North-South axis), nonlinear
optics are numerically optimized for CHESS-U using 76
independently-controlled sextupoles with an implementation of J.
Bengtsson's resonance driving term (RDT) formalism
\cite{SLS:TME:TA1997:0009} in Bmad. As there is no symmetry, all
sextupoles are allowed to vary separately.

An unusual design feature of the CHESS-U achromat is a lack of
sextupoles in the new sextant. The peak dispersion in the DBAs is
around 20~cm, whereas the average dispersion in the remaining 5/6 of
CESR is around 1~m; the betas in the new achromats are also
suppressed compared to the rest of the machine. As such, any
sextupoles in the new sextant have little leverage on the
chromaticity. Optimizations showed the inclusion of harmonic
sextupoles in the new DBA cells did not significantly improve the
nonlinear optics, and were therefore omitted.

The new DQ magnets have a significant $b_2$ multipole, contributing
roughly $\mp 0.5$ to the horizontal and vertical chromaticity,
respectively. This was compensated during the RDT reduction
optimization.

%

%% file: s3_lattice_characterization.tex
\subsection{Performance}

\begin{figure}
\centering
    \includegraphics[width=0.45\textwidth]{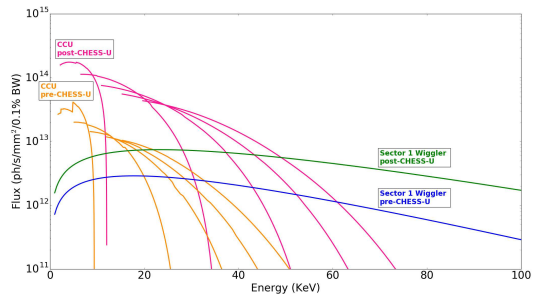}
    \caption{SPECTRA modeling of flux through a 1~mm$^2$ pinhole at 20 meters,
    before and after CHESS-U upgrade \cite{CHESSU:ScienceCase}. ``F-line wiggler''
    refers to the 24-pole wiggler for Sector 1.}
    \label{fig:spectra}
\end{figure}

Four of the five sectors with CHESS IDs will have canted 1.5~m CHESS
Compact Undulators (CCUs) \cite{SRI12:032004, SPIE:951205,
IEEE-Trans:4100504}. Sector 1 will use the existing 24-pole F-line
wiggler \cite{RSI63:305-308}. Anticipated pinhole flux was simulated
with SPECTRA \cite{JSR:1221-8}, shown in Fig. \ref{fig:spectra}.

\subsection{Error Tolerance}

    Error tolerance for CHESS-U was assessed using the tools and
    low-emittance tuning methodology developed for the CesrTA program
    \cite{PRSTAB17:044003}.
    The optics correction procedure used here is identical to that in CesrTA:

    \begin{enumerate}
        \item Measure closed orbit; correct using horizontal and
        vertical steering magnets.
        \item Measure betatron phase advance, local coupling
        \cite{PRSTAB3:102801}, and horizontal dispersion; correct
        using quadrupoles and skew quadrupoles.
        \item Measure orbit, local coupling, and vertical dispersion;
        correct using vertical steerings and skew quadrupoles.
    \end{enumerate}

    Though the new DQs have trim windings, the
    correction simulation did not allow for the DQs to vary.
    Error amplitudes are listed in Tables \ref{tab:ringma_errors} and \ref{tab:bpm_errors},
    and represent the best estimate of anticipated alignment and BPM errors.

\begin{figure}
    \centering
        \includegraphics[width=0.45\textwidth]{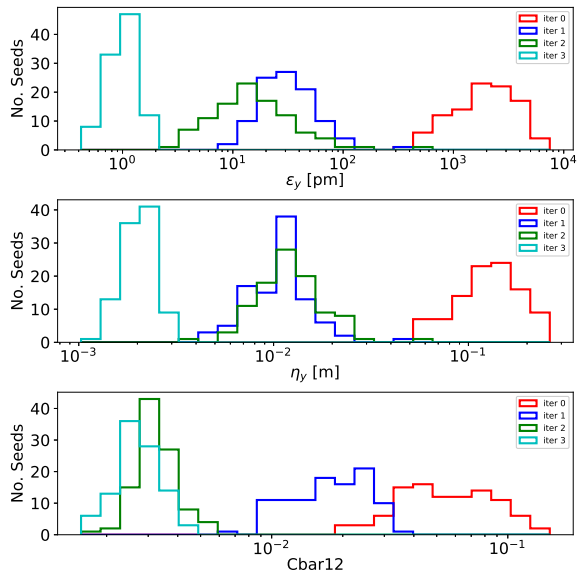}
        \caption{Statistical analysis of 100 misaligned and corrected lattices.
        Iter0 corresponds to the misaligned lattice before correction; the following
        three iterations correspond to the steps listed above. $95^{th}$-percentile
        $\epsilon_y = 4.1$~pm after the final correction.}
        \label{fig:ringma}
    \end{figure}

\begin{figure}
    \centering
        \includegraphics[width=0.45\textwidth]{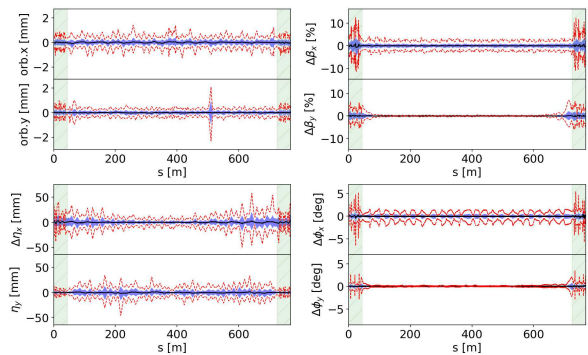}
        \caption{Statistics of machine optics distortion as a function of
        longitudinal position for 95 out of 100 misaligned and
        corrected lattices (representing 95th-percentile anticipated results).
        Black curve indicates the average over 100 seeds,
        blue shading indicates the standard deviation, and the red dashed lines
        indicate the maximum and minimum excursion at any given point in the
        lattice. The green hatched region indicates the region with new CHESS-U achromats,
        where betas and dispersion are suppressed compared to the rest of the ring.}
        \label{fig:ringma_twiss}
    \end{figure}

    A summary of the misalignment study is shown in Fig.
    \ref{fig:ringma}. The statistical impact of misalignments and
    corrections on the Twiss parameters is shown in
    Fig. \ref{fig:ringma_twiss}. The $95^{th}$-percentile vertical
    emittance due to optics errors after correction is 4.1~pm.

\subsection{Dynamic and Momentum Aperture}

    Dynamic and momentum aperture were evaluated via frequency map
    analysis \cite{LASKAR1993257}. Particles were tracked for a total of 2048 turns at
    each initial amplitude; dynamic aperture plots were normalized to
    beam sigmas, assuming 1\% emittance coupling.

    Systematic multipoles for dipoles,
    quadrupoles, and sextupoles were
    determined via field modeling or measurement, and are
    listed in Table \ref{tab:systematic_multipoles}.
    Due to the compact nature of the CCU magnet arrays, the field
    roll-off is significant: vertical field is reduced 10\% at a
    10~mm horizontal displacement. Systematic multipoles and the field roll-off and
    measured field integrals for CCUs were included in some tracking simulations
    \cite{Wang:IPAC2016-WEPOW053}, denoted in the relevant figure captions.

\begin{figure*}[tbh]
    \centering
        \begin{subfigure}[t]{\textwidth}
            \includegraphics[width=0.45\textwidth]{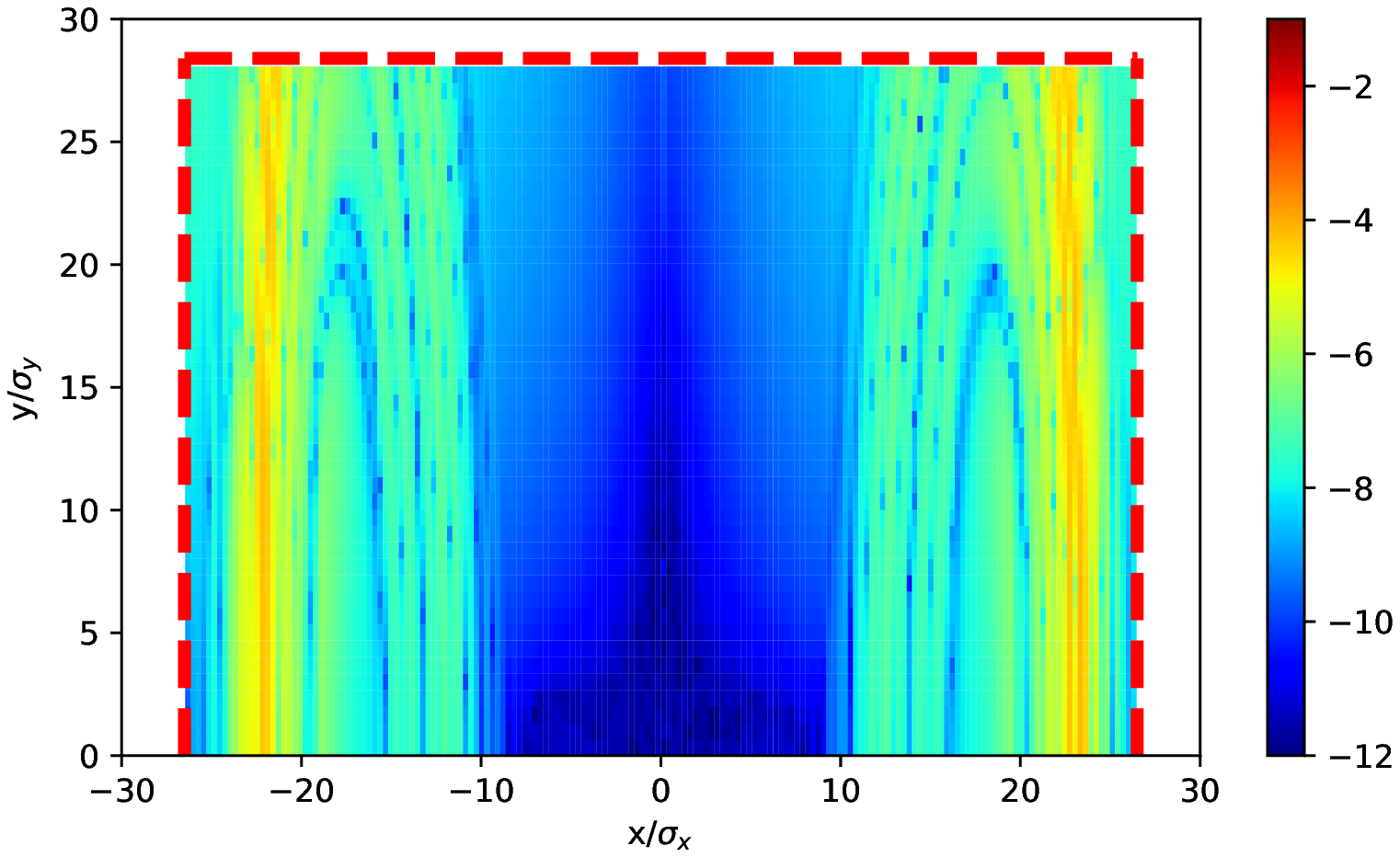}
            \includegraphics[width=0.45\textwidth]{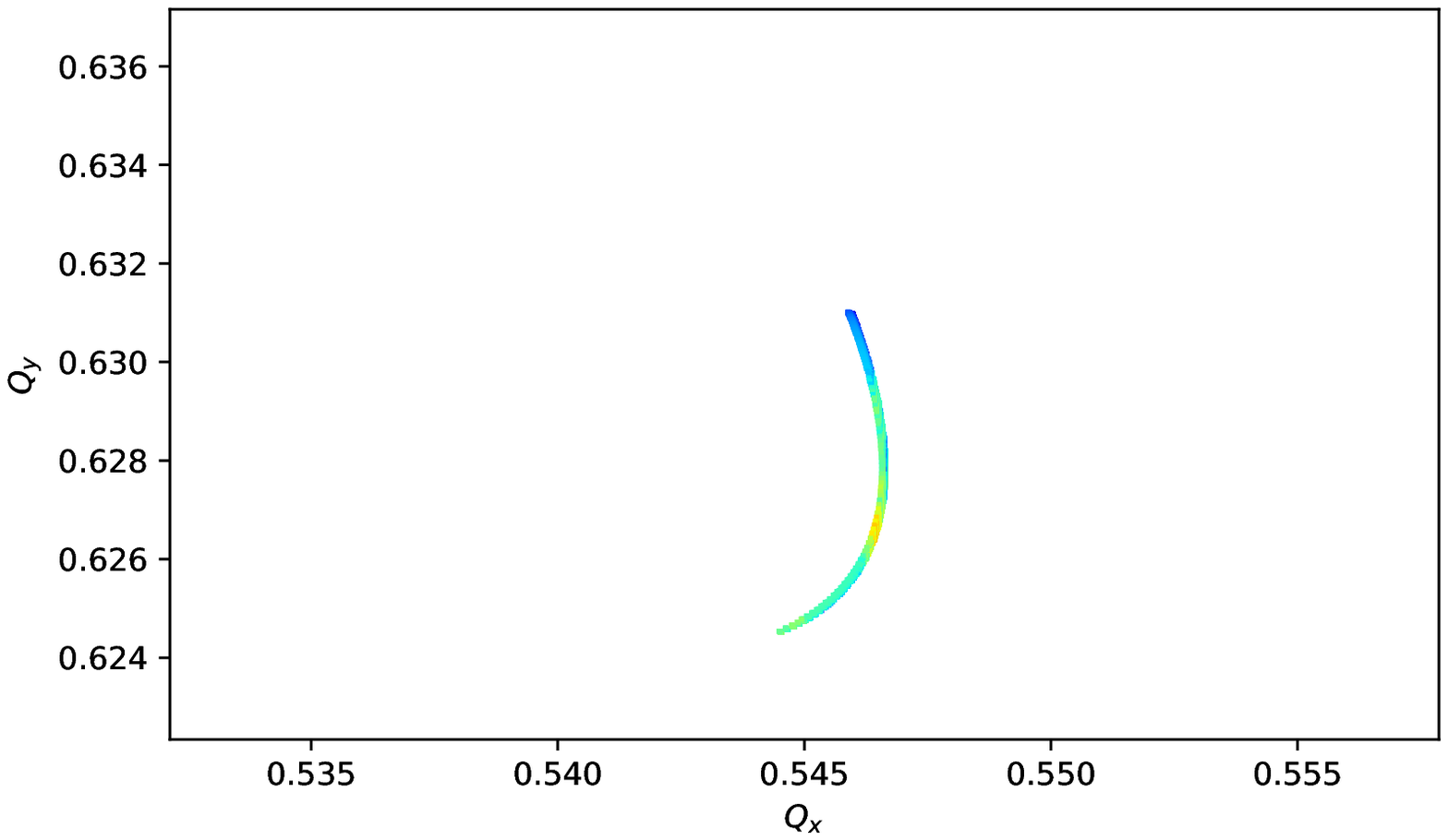}
            \caption{Ideal lattice.}\label{fig:freq_map_da_ideal}
        \end{subfigure}
        \begin{subfigure}[t]{\textwidth}
            \includegraphics[width=0.45\textwidth]{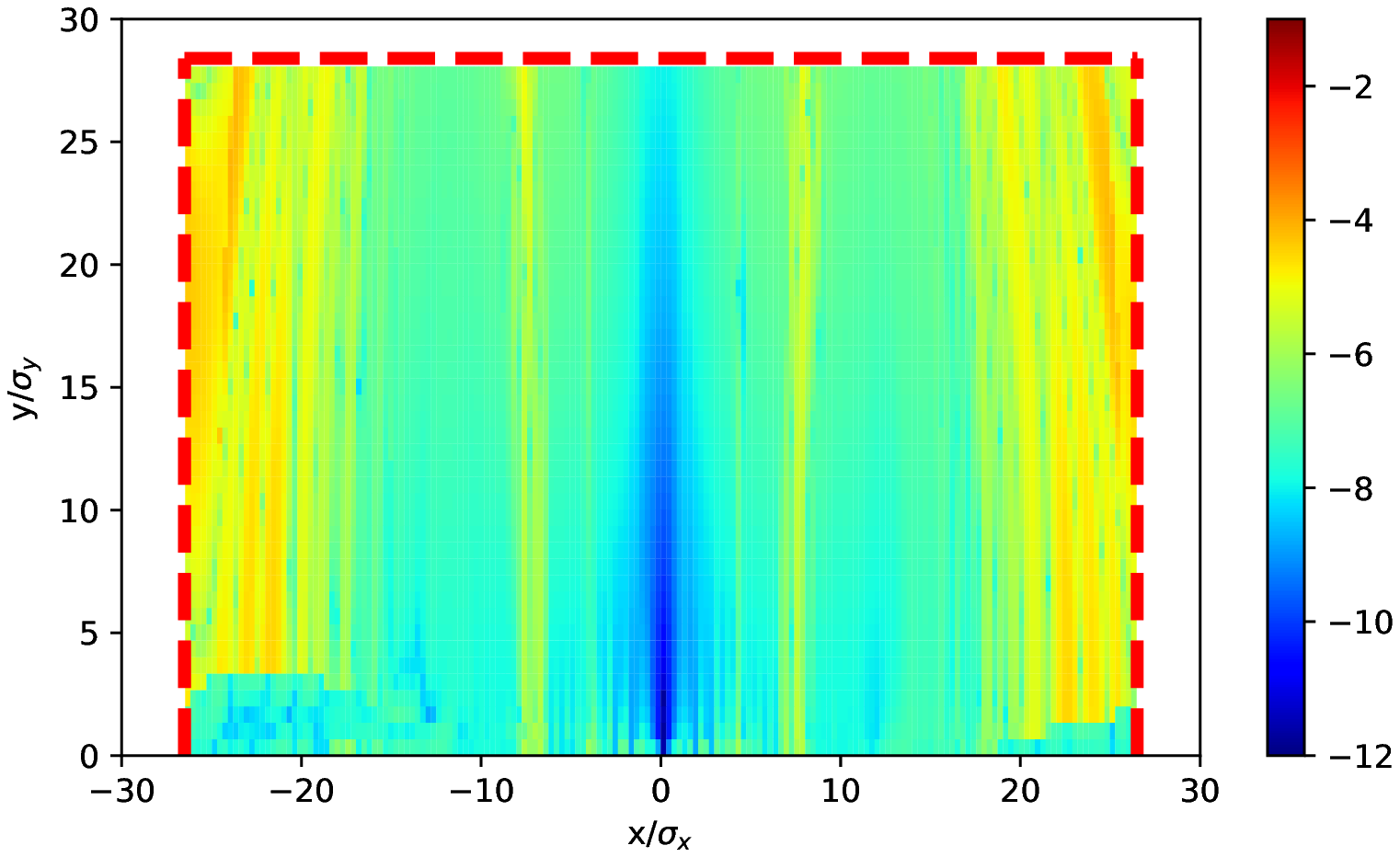}
            \includegraphics[width=0.45\textwidth]{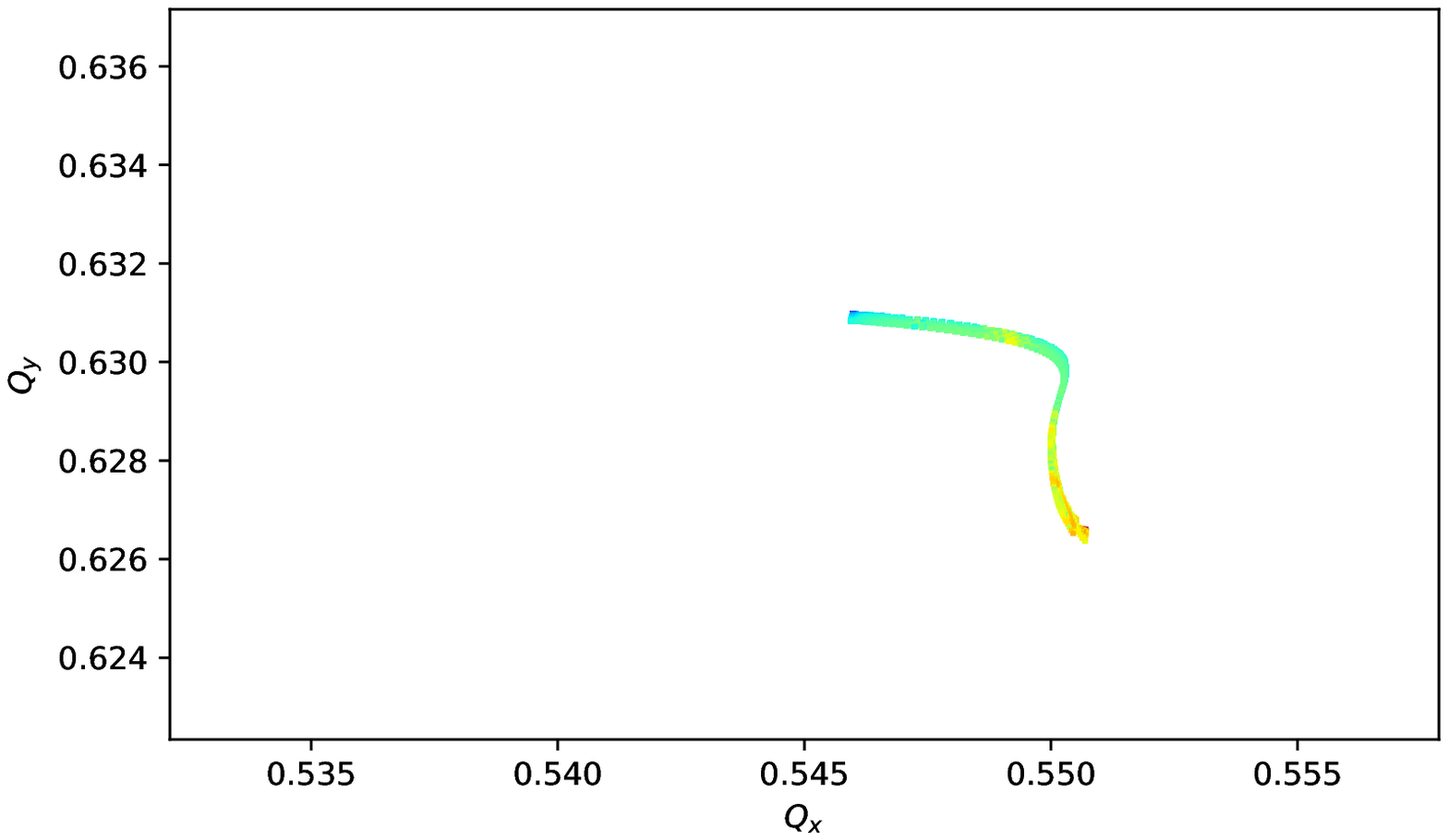}
            \caption{With systematic multipoles and realistic undulator field tracking.}
            \label{fig:freq_map_da_mult_all}
        \end{subfigure}
        \begin{subfigure}[t]{\textwidth}
            \includegraphics[width=0.45\textwidth]{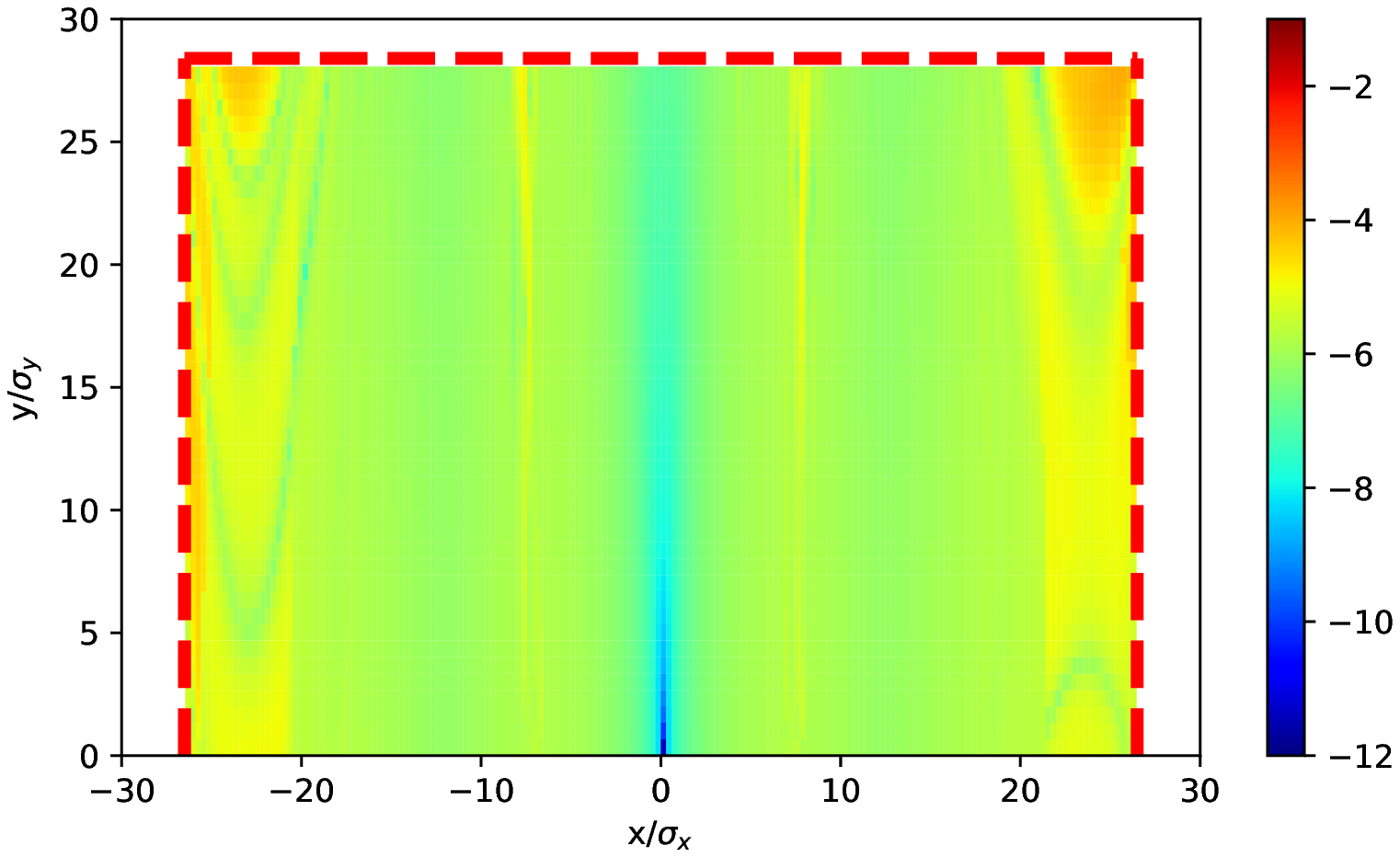}
            \includegraphics[width=0.45\textwidth]{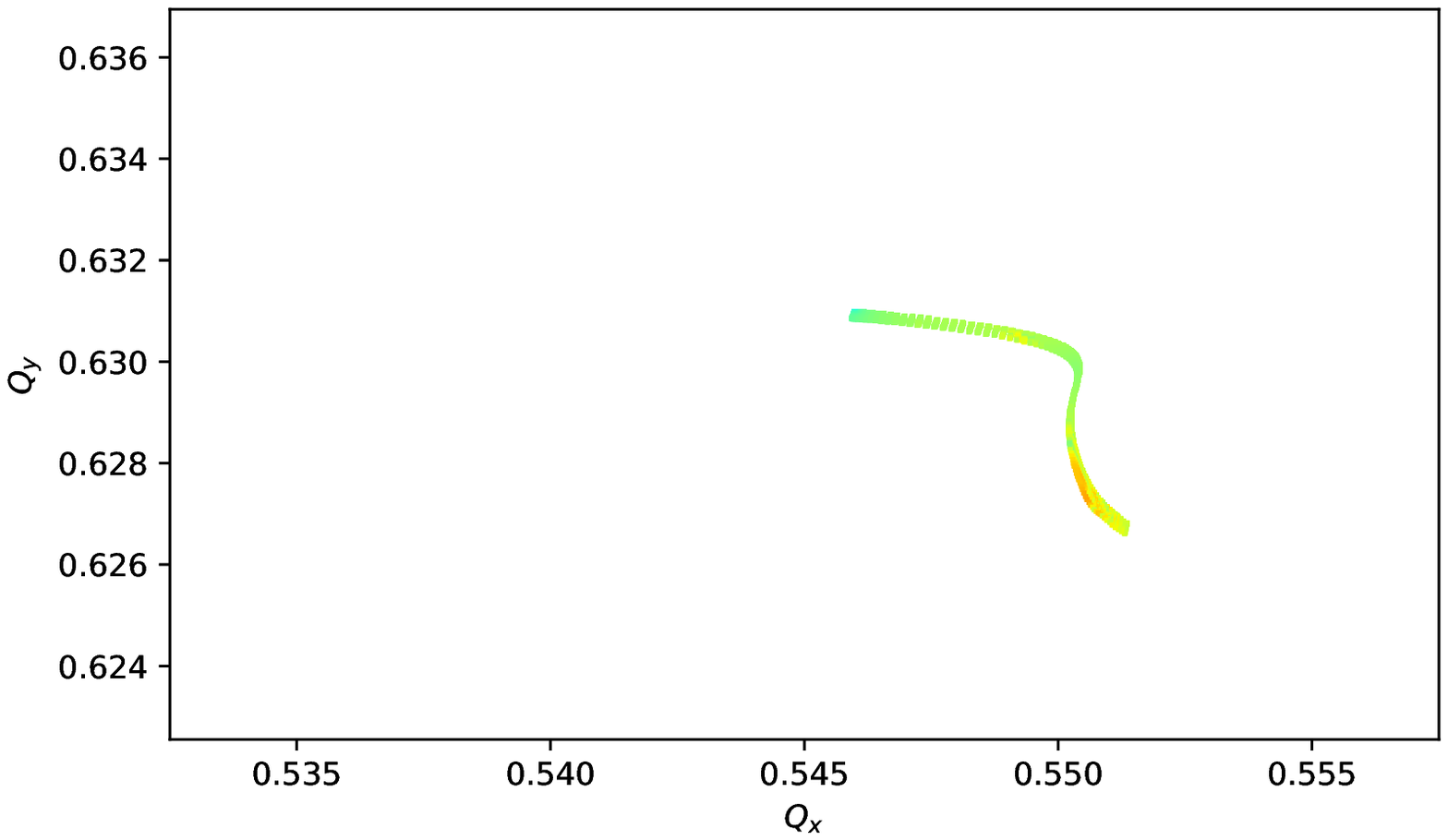}
            \caption{With systematic multipoles, realistic undulator field tracking,
                        and misalignments and corrections ($95^{th}$-percentile seed).}\label{fig:freq_map_da_ma}
        \end{subfigure}
        \caption{Dynamic aperture via the frequency map (left column), and its associated footprint in the tune plane (right column).
        Color scale indicates tune diffusion, defined in Eqn. \ref{eqn:tune_diffusion}. Red dashed line indicates
        the minimum projected aperture.}
        \label{fig:freq_map_da}
    \end{figure*}

\begin{figure*}[tbh]
    \centering
        \begin{subfigure}[t]{\textwidth}
            \includegraphics[width=0.45\textwidth]{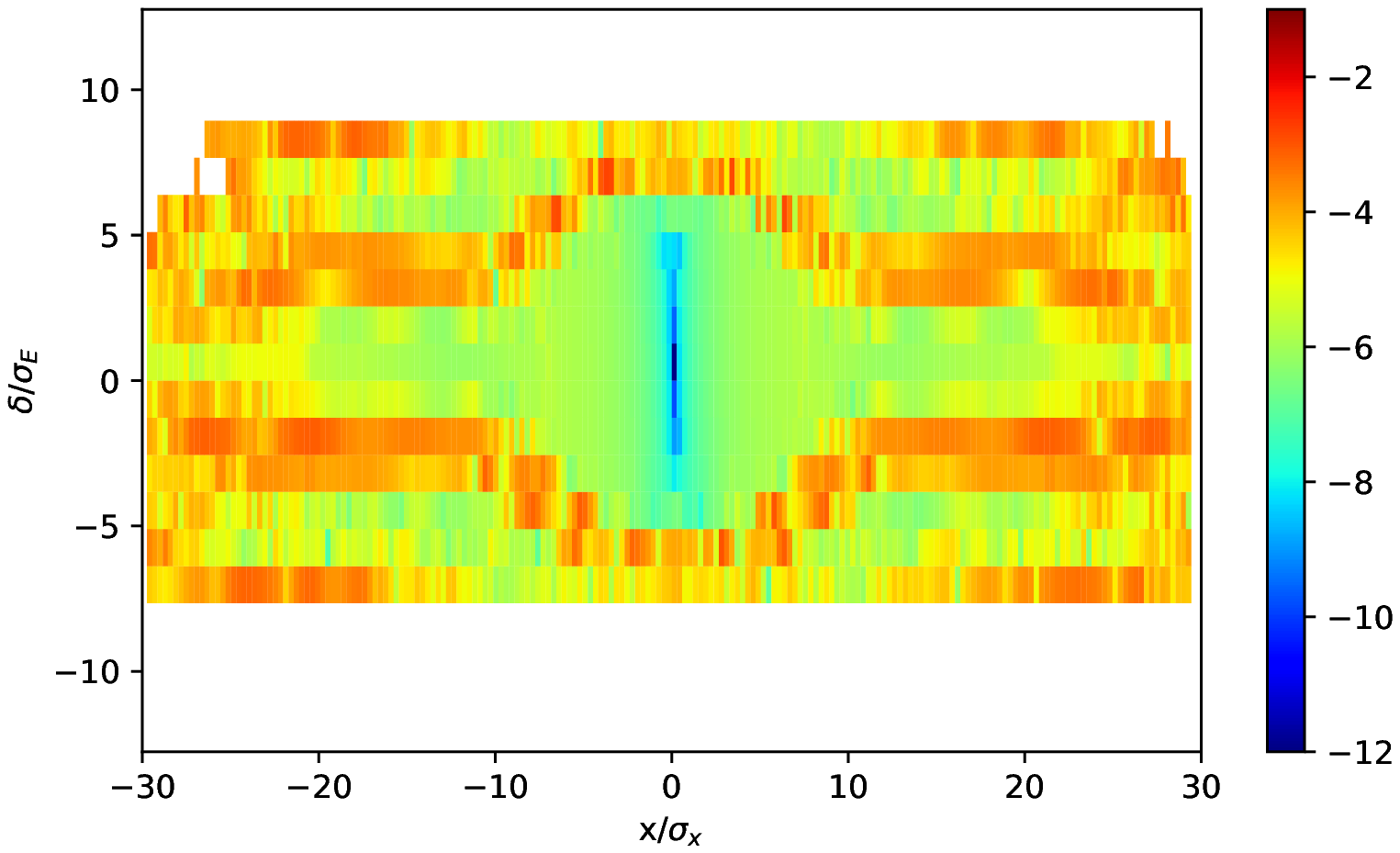}
            \includegraphics[width=0.45\textwidth]{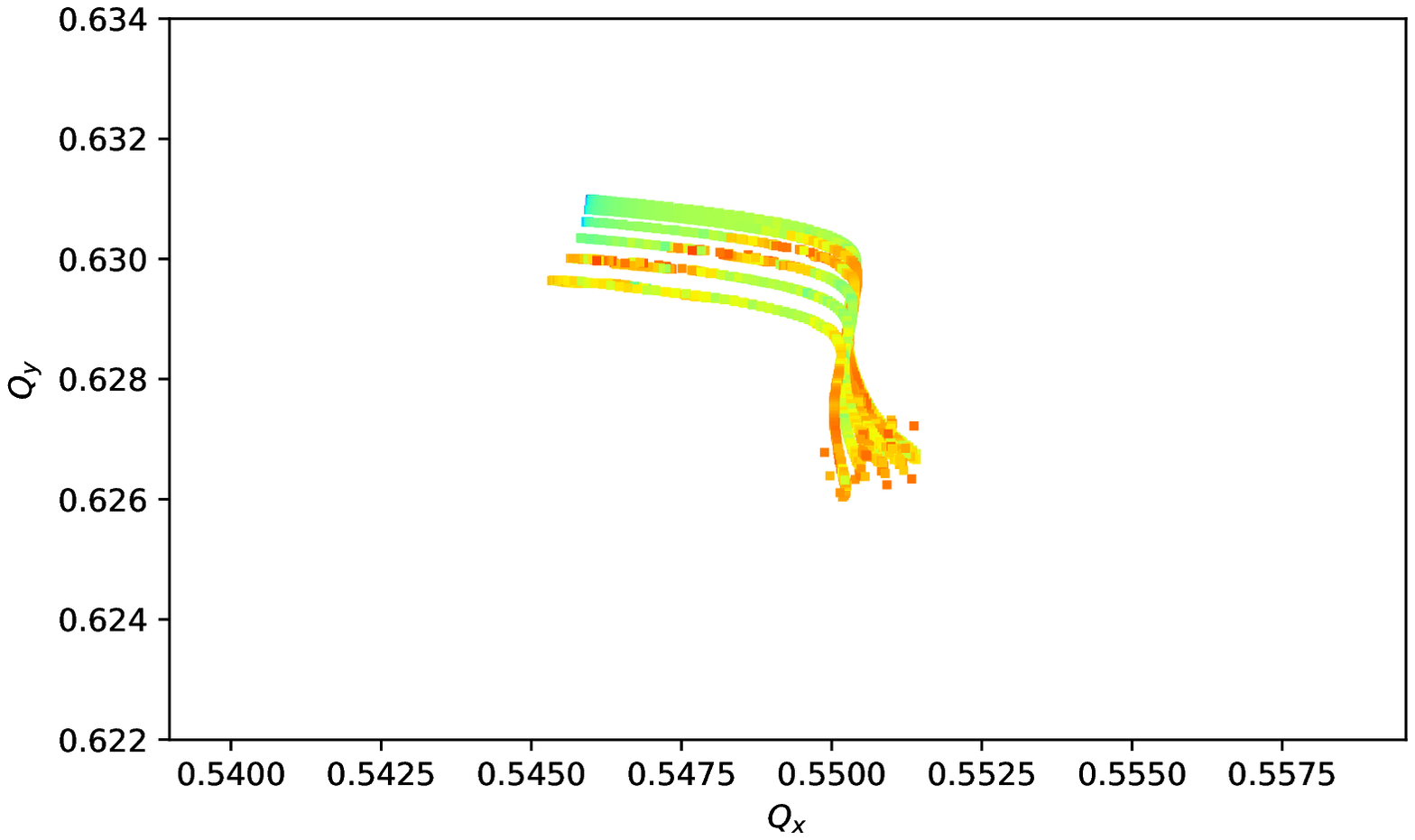}
        \end{subfigure}
        \caption{Momentum aperture via the frequency map and its associated footprint
        in the tune plane, with systematic multipoles, realistic undulator field tracking,
        and misalignments and corrections ($95^{th}$-percentile seed).
        Color scale indicates tune diffusion, defined in Eqn. \ref{eqn:tune_diffusion}.}
        \label{fig:freq_map_ma}
    \end{figure*}

    Dynamic aperture and its footprint in the tune plane are shown in Fig.
    \ref{fig:freq_map_da} for combinations of systematic multipoles, insertion
    devices, and misalignment/correction; corresponding momentum
    aperture with systematic multipoles, IDs, and misalignment/correction
     are shown in Fig. \ref{fig:freq_map_ma}. The color scale in
     these plots is the tune diffusion, given by

     \begin{eqnarray}\label{eqn:tune_diffusion}
     dQ = \log\sqrt{\Delta Q_x^2 + \Delta Q_y^2}
     \end{eqnarray}

\subsection{Injection Simulation}

    The injection scheme for CESR relies on accumulation from a 60~Hz
    synchrotron booster ring. Injection efficiency was simulated via
    multi-particle tracking. 200 macroparticles were launched with a
    phase space matching the beam at the end of the synchrotron
    booster extraction line. Particles which survive for 400 turns
    were considered captured. Details are found in
    \cite{Wang:IPAC2016-WEPOW053}.

\begin{figure}
    \centering
    \includegraphics[width=0.45\textwidth]{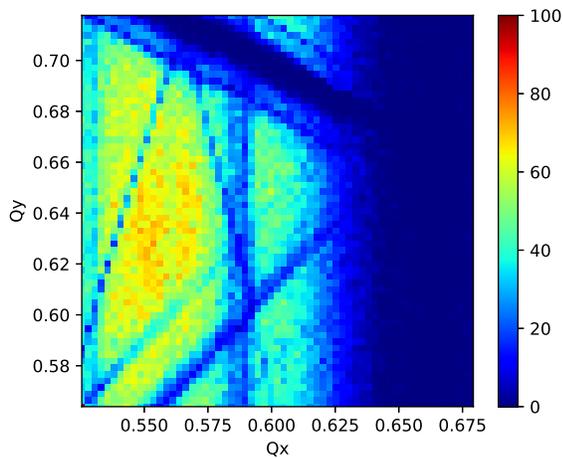}
    \caption{Injection tune scan simulation with systematic multipoles,
    realistic undulator field tracking, and misalignments and corrections
    ($95^{th}$-percentile seed). Color scale indicates capture efficiency
    in percent.}
    \label{fig:injtrack}
    \end{figure}

    Results of the injection tracking simulations are shown in Fig.
    \ref{fig:injtrack}. It should be noted that the injection efficiency is not
    expected to reach 100\% as there are two 10~mm full-aperture vertical collimators
    used to mask against particle losses on the small-gap permanent-magnet IDs.

%% file: s4_magnet_design.tex

Most constraints on magnet design are more or less standard. The
specific design impacts on the CHESS-U magnets are discussed here.
Details are available in \cite{CHESSU-ExtRev:Magnets}.

\subsection{Combined Function Dipole-Quadrupoles}

\begin{figure}
\centering
    \includegraphics[width=0.3\textwidth]{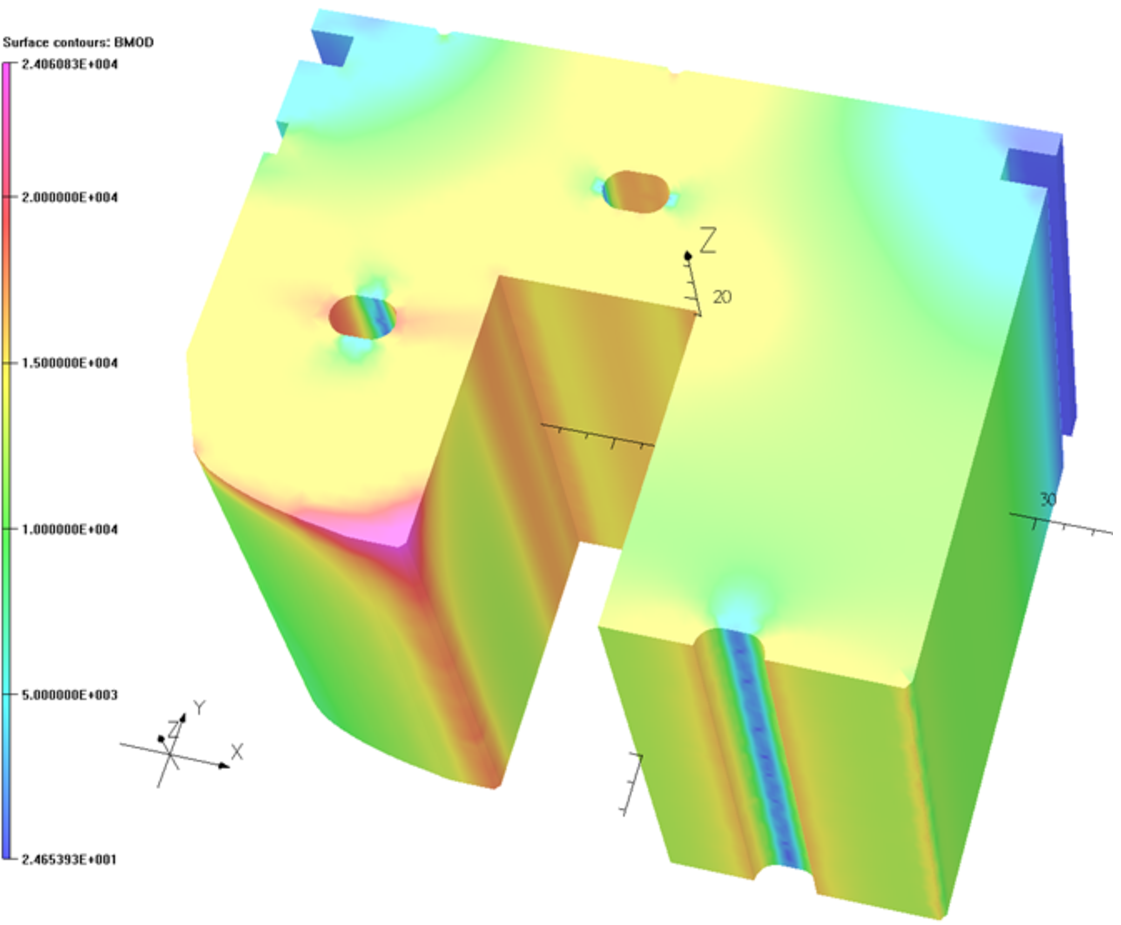}\\
    \vspace{0.5 cm}
    \includegraphics[width=0.35\textwidth]{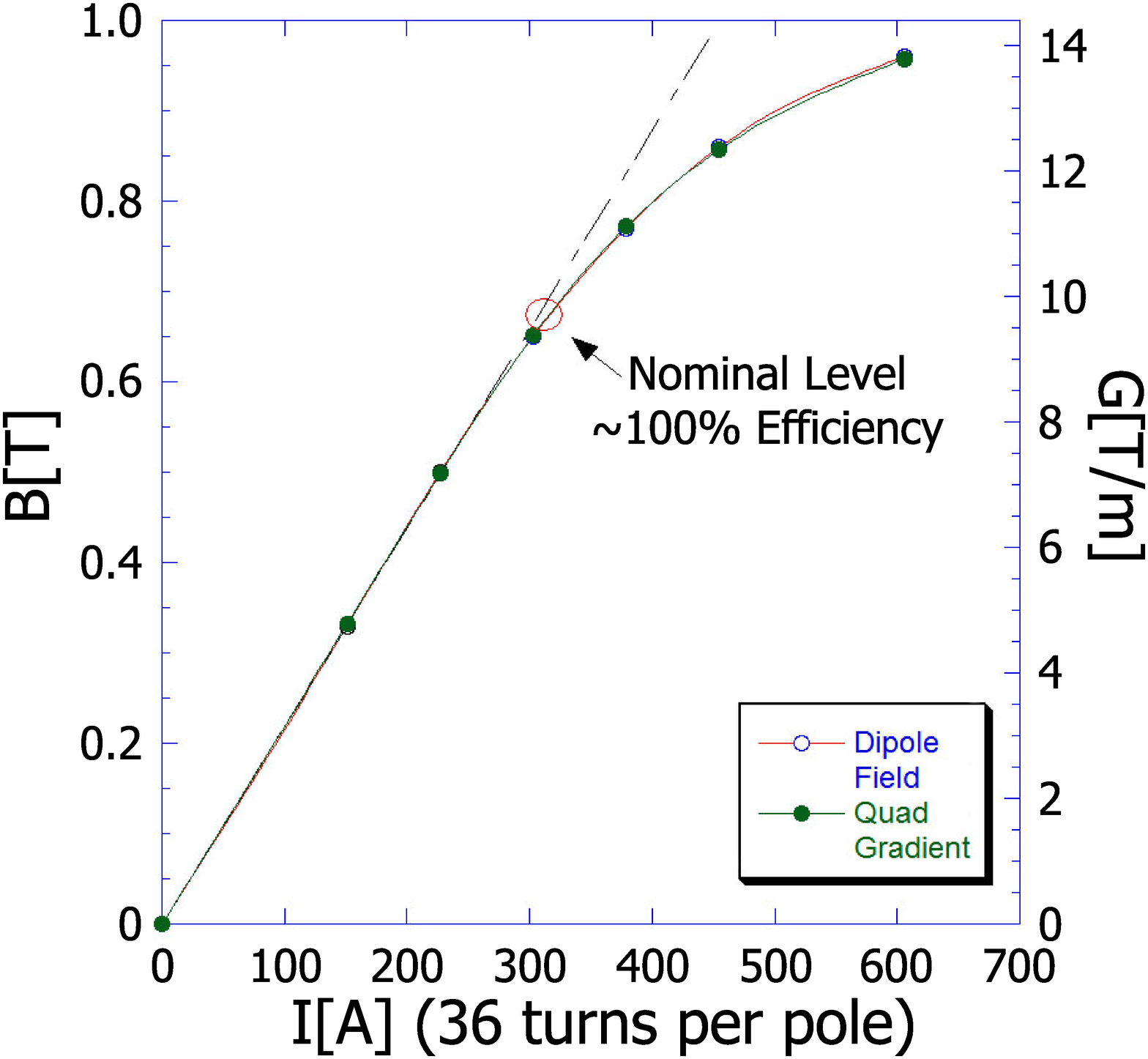}
    \caption{Top: Magnetic field in one half of a DQ iron.
    Peak field in the iron is approximately 2.3~T.
    Bottom: DQ excitation curve.}
    \label{fig:dq}
\end{figure}

Space constraints and emittance reduction motivated implementing DQs
in the new DBA cells rather than conventional dipoles. The DQs have
a magnetic length of 2.35~m, bend radius of 31.4~m, and gradient
-8.77~T/m. The full-gap aperture along the design trajectory is
38.4~mm. The 3D OPERA model and excitation curve for the DQ magnet
are shown in Fig. \ref{fig:dq}. Although the magnet yoke is
rectangular in geometry, the pole pieces follow a sector dipole
trajectory.

\subsection{Quadrupoles}

\begin{figure}
\centering
    \includegraphics[width=0.3\textwidth]{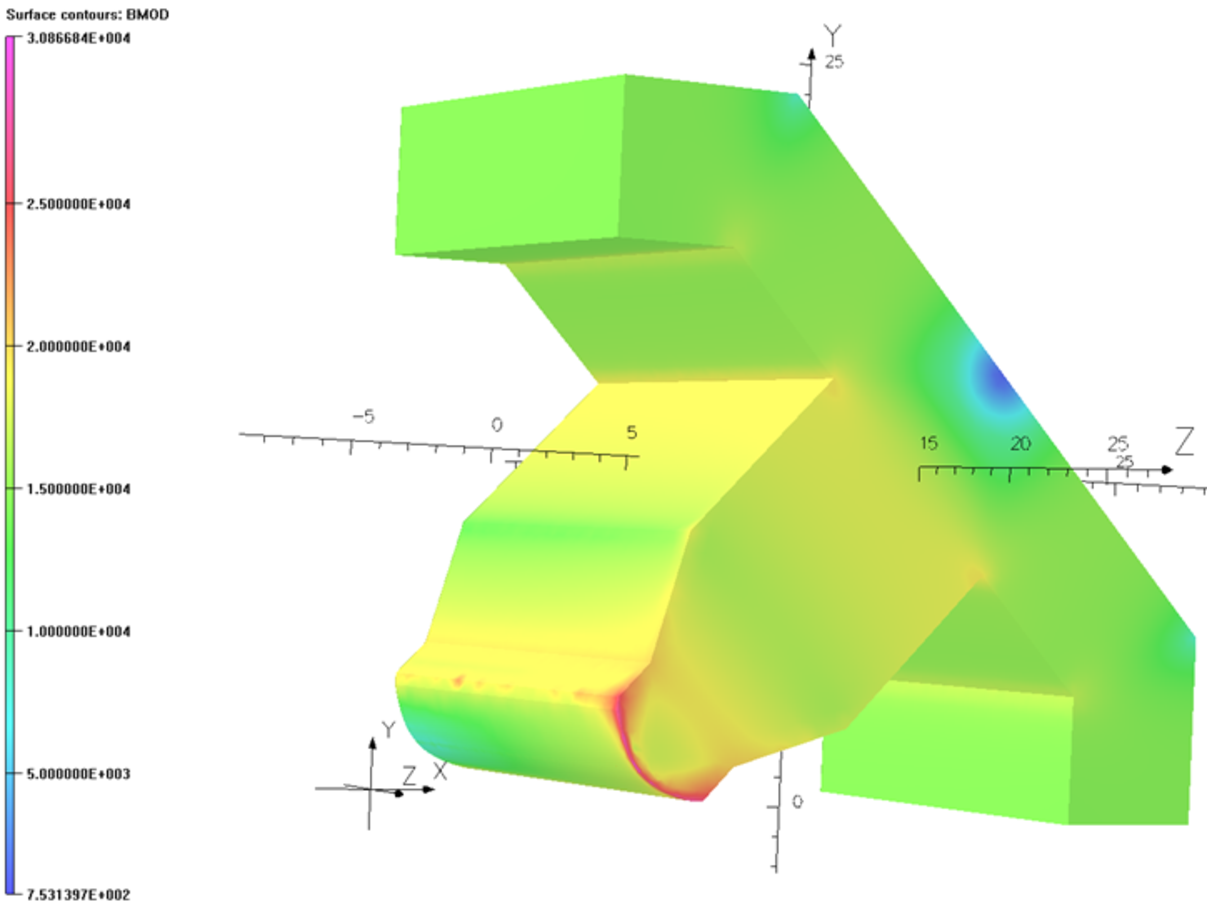}\\
    \vspace{0.5 cm}
    \includegraphics[width=0.35\textwidth]{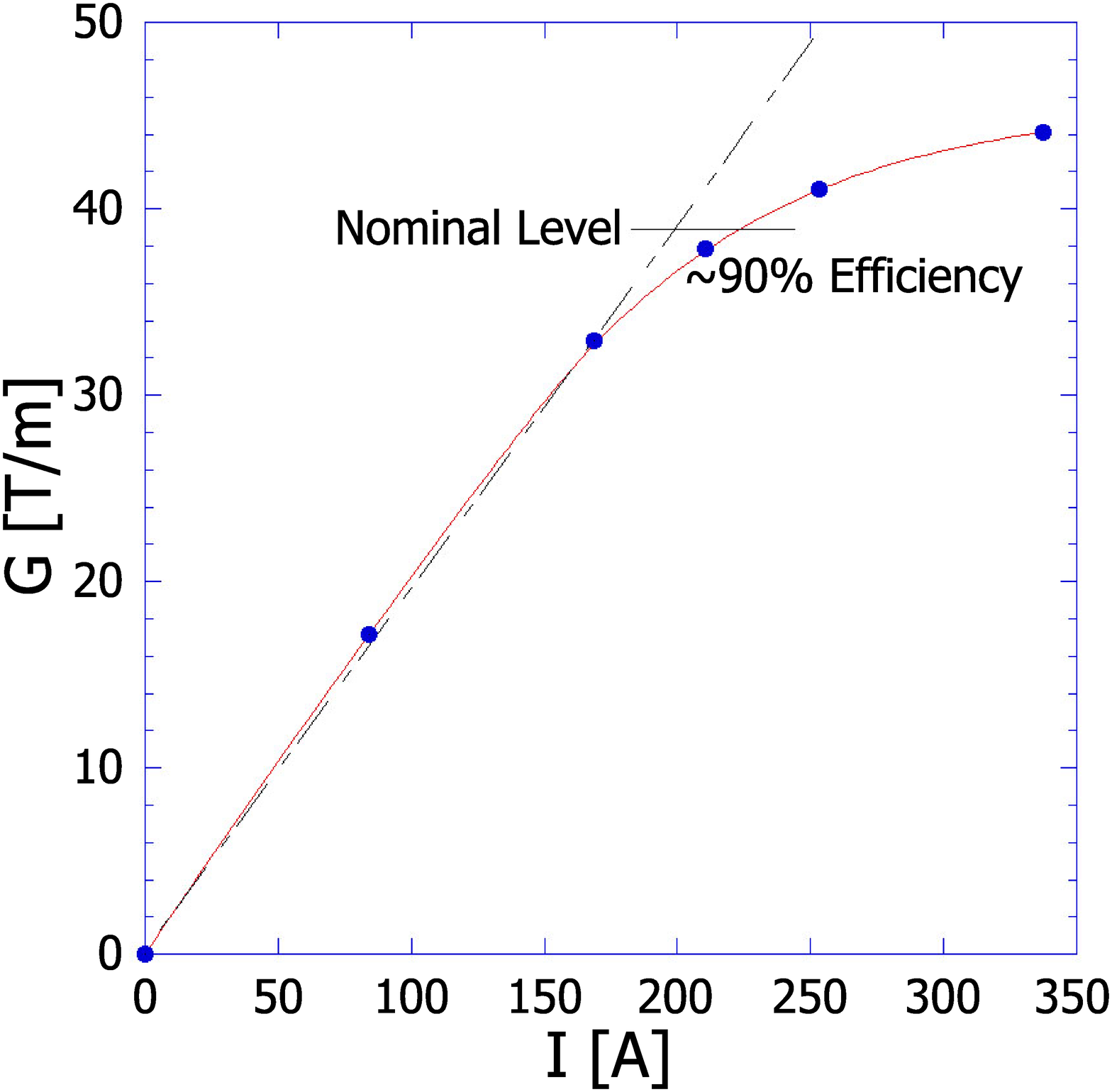}
    \caption{Top: Magnetic field in one quarter of a non-extraction
    quadrupole iron. Peak field in the iron is approximately 2.3~T.
    Bottom: Quadrupole excitation curve. }
    \label{fig:quads}
\end{figure}

\begin{figure}
\centering
    \includegraphics[width=0.4\textwidth]{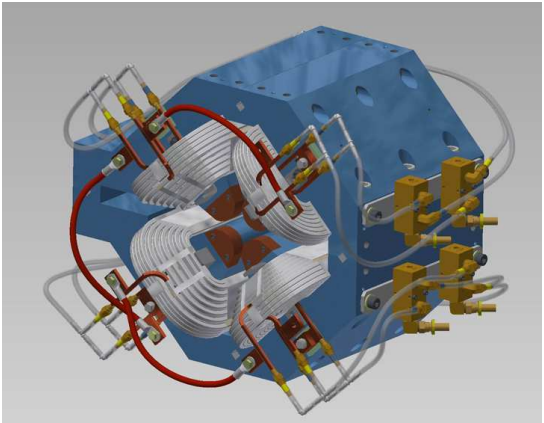}
    \caption{Extraction quadrupole model. }
    \label{fig:quad_extraction}
\end{figure}

All achromat quadrupoles are horizontally-focusing, as the vertical
focusing is done by the DQs. There are two families of quadrupoles,
with magnetic lengths 0.400~m and 0.362~m, both with the same pole
profile; the primary distinctions are length and the extension of
the iron to allow for x-ray extraction on one of the two families.
Both quadrupole families have maximum gradient 39~T/m, with a bore
radius of 23~mm. Fig. \ref{fig:quads} shows the magnetic field in
one quarter of a non-extraction quadrupole and the excitation curve;
an extraction quad model is shown in Fig. \ref{fig:quad_extraction}.

%

\subsection{CHESS Compact Undulators}
\begin{figure}
\centering
    \begin{subfigure}{0.45\textwidth}
    \includegraphics[width=\textwidth]{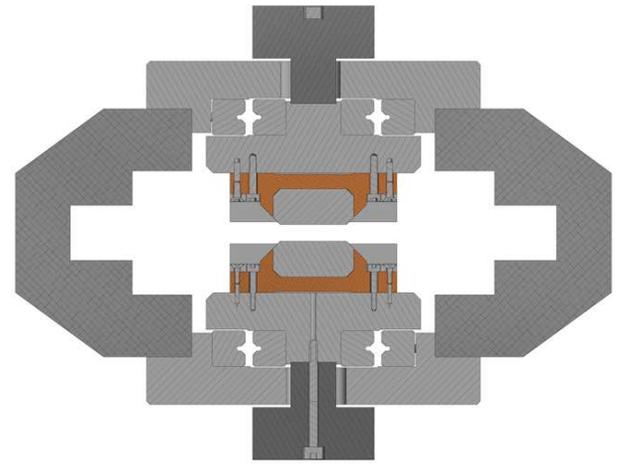}
    \caption{End-on view. }
    \label{fig:ccu_slice}
    \vspace*{10mm}
    \end{subfigure}
    \begin{subfigure}{0.45\textwidth}
    \includegraphics[width=\textwidth]{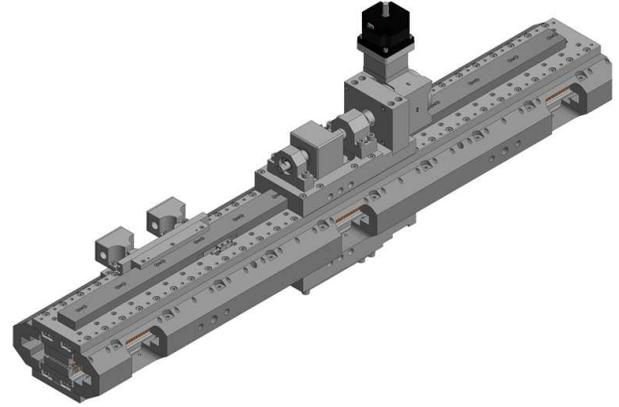}
    \caption{Perspective view. }
    \label{fig:ccu_perspective}
    \end{subfigure}
    \caption{CAD model of the Cornell Compact Undulator (CCU).}
    \label{fig:ccu}
\end{figure}

Four of the five new sectors will be equipped with canted pairs of
CHESS Compact Undulators (CCUs). The design for the CCUs is detailed
elsewhere \cite{SRI12:032004, SPIE:951205, IEEE-Trans:4100504}. An
end-view of the CCU profile is shown in Fig. \ref{fig:ccu_slice};
perspective CAD drawing of CCU assembly is shown in Fig.
\ref{fig:ccu_perspective}. Parameters for the CCUs are summarized in
Table \ref{tab:ccu_params}.

\begin{table}[htb]
 \begin{centering}
     \begin{tabular}{cccc}
         \toprule
         \textbf{Parameter}    & \textbf{Value}\\
         \colrule
         Length  & 1.477~m  \\
         Period  &  28.4~mm \\
         Peak field & 0.952~T \\
         Pole gap  & 7.0~mm \\
         First harmonic &  2.9~keV \\
         \botrule
     \end{tabular}
     \caption {Parameters for the CHESS Compact Undulators.}
     \label{tab:ccu_params}
 \end{centering}
 \end{table}

The CCUs utilize a variable-phase design rather than variable-gap in
order to minimize the supporting infrastructure. As implemented in
CESR the CCUs have a 7.0~mm out-of-vacuum vertical aperture between
poles, mandating the vacuum chamber to fit inside this aperture. The
associated vacuum chamber for the CCUs is described in Section
\ref{sec:vacuum_chambers}.

CHESS has operated with two 1.5~m CCUs in a canted straight since
2014, with two counter-rotating beams. Total beam current has
routinely exceeded the CHESS-U target 200~mA in single-beam positron
tests at the present CHESS beam energy of 5.3~GeV.

\subsection{Power Distribution}

For the new CESR south arc, the magnet power distribution was chosen
to be be compatible with existing infrastructure. There are two main
pieces to this: the CESR dipole circuit and the CESR quadrupole bus
system. The controls for both are derived from a unique computer bus
known as the Xbus \cite{IEEETrans:4330007, IPAC12:THPPR015}. The
Xbus handles all digital and analog controls, read back signals, and
interlocks associated with the operation of the magnet power
supplies. All magnets can be controlled individually or collectively
as needed from the central control room.

The CESR Dipole circuit is a series circuit of approximately 120
bend magnets driven by two 0.5~MW Transrex power supplies at 500~V
and 740~A to obtain 6~GeV. The series chain is made up of normal
bend, soft bend, and hard bend dipole magnets. The new DQ magnets
will be part of this existing circuit.

The operation of CHESS-U requires five quadrupole buses: East, South
East, South Center, South West, and West. The SE, SC, and SW are new
buses installed expressly for the south arc installation. The West
and East quad buses are preexisting. Each quad bus is driven by an
80~V 1125~A EMHP power supply. These buses supply a fixed 70~V
primary voltage to a large number of switching power supplies
(DC-to-DC converters). These power supplies excite all quadrupole,
sextupole, steering, skew quadrupole, skew sextupole, and octupole
magnets. Through the Xbus, each switching power supply can be
controlled independently, giving flexibility for tuning optics. The
current draw depends on the optics, typically scaling with beam
energy.

\subsection{Support and Alignment}

\begin{figure}
\centering
    \includegraphics[width=0.45\textwidth]{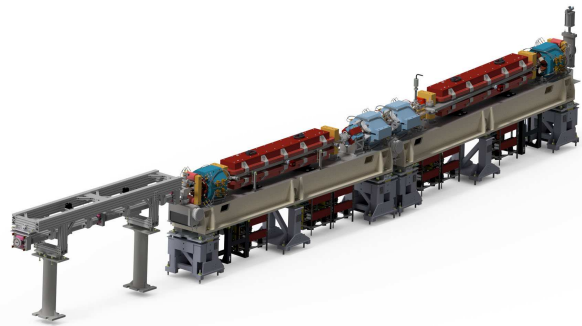}
    \caption{Rendering of one sector for CHESS-U from the radial-outside.
    Beam travels clockwise (right-to-left).}
    \label{fig:sector_rendering}
\end{figure}

Each new CHESS-U cell was split into three sections: an upstream
girder, a downstream girder, and an insertion device straight. The
two girders support all achromat magnets, each holding two
quadrupoles and one DQ, plus corrector magnets. The insertion device
straight is suspended from above in order to allow access to the
upstream sector's front end. One sector is illustrated in Fig.
\ref{fig:sector_rendering}.

Magnetic alignment of the DQ and quadrupoles is achieved using a
combination of vibrating wire \cite{nima-399:185} and Hall probe.
The procedure is as follows:

\begin{enumerate}
    \item Establish vibrating wire (VW) position with respect to
    girder fiducials
    \item Establish Hall probe position with respect to VW
    \item Place wire on beam axis
    \item Align Q1 magnetic axis with VW
    \item Align Q2 magnetic axis with VW
    \item Align DQ using Hall probe
\end{enumerate}

\begin{figure}
\centering
    \includegraphics[width=0.3375\textwidth, valign=c]{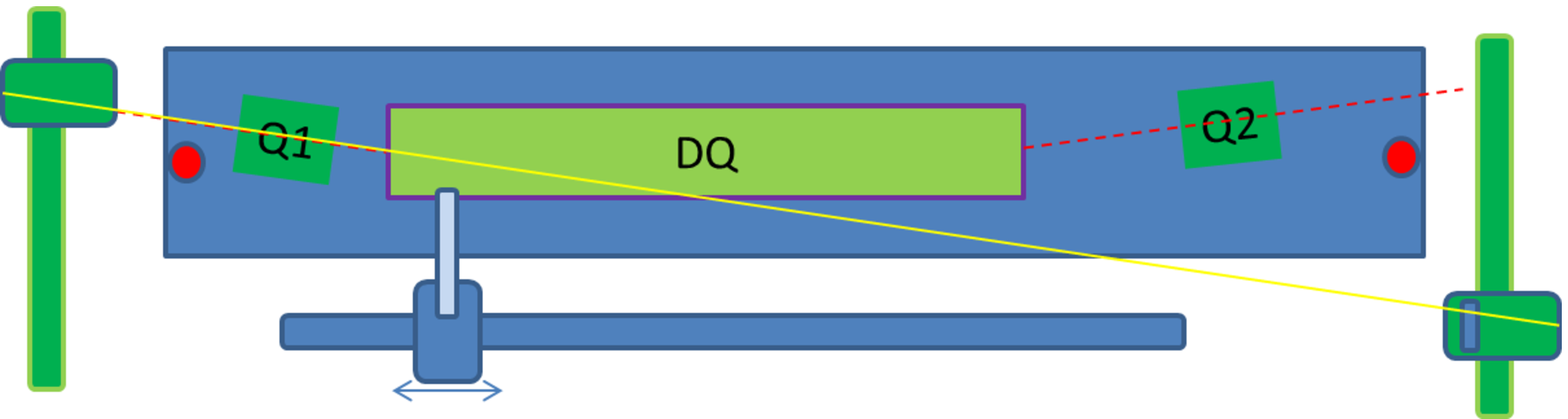}
    \includegraphics[width=0.1125\textwidth, valign=c]{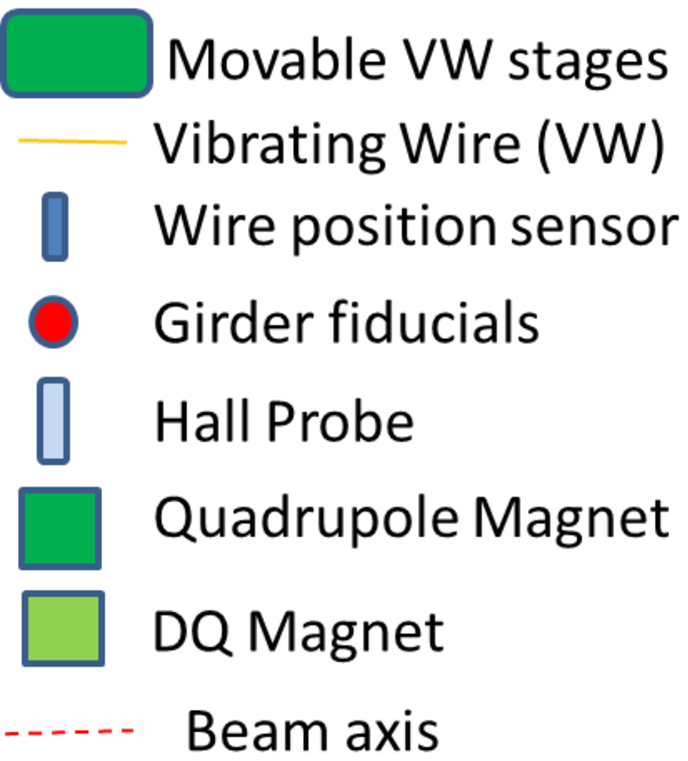}
    \caption{Setup for magnetic centering of DQ and quadrupoles on a girder. \cite{CHESSU-ExtRev:Magnets}}
    \label{fig:magnetic_alignment}
\end{figure}

The setup used for this alignment procedure is diagrammed in Fig.
\ref{fig:magnetic_alignment}. Using this method, the magnetic
centers of quadrupoles and the DQ on one girder are aligned to
within 32~$\mu$m. 

%

%% file: s5_vacuum_design.tex
\subsection{Design Considerations}

As a part of particle beam transport system, the CHESS-U vacuum
chambers need to provide adequate physical beam aperture, while
maintaining sufficient clearance to all CHESS-U magnets. The vacuum
pumping system is designed to achieve and to maintain ultra-high
vacuum (UHV) conditions, with an average vacuum pressure $<$ 1~nTorr
 with stored 200~mA beam at 6.0~GeV. Additionally, all vacuum
chambers must be designed to manage heating from synchrotron
radiation (SR) generated by the bending magnets, with a minimum
factor of safety (FOS) $>$2. The vacuum chambers will also house a
suite of BPMs for beam instrumentation.

\subsection{Vacuum Chambers}\label{sec:vacuum_chambers}

\begin{figure*}
\centering
    \includegraphics[width=\textwidth]{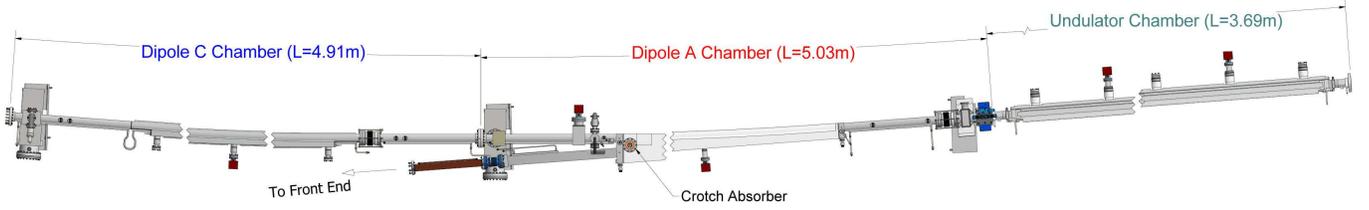}
    \caption{A typical CHESS-U Achromat Cell vacuum chamber string is consisted of three long vacuum chambers.}
    \label{fig:vac-achromat}
\end{figure*}

All CHESS-U vacuum chambers are constructed from type 6061 and 6063
aluminum alloys.  Similar to the modular magnet design concept, the
CHESS-U vacuum chambers conform to the periodic achromat cells.
Owing to the limited longitudinal spaces between the magnets, the
vacuum string for each achromat is made into three flanged chambers,
as shown in Fig. \ref{fig:vac-achromat}. More specifically, each
achromat is comprised of two long dipole chambers, a 4.91~m long
downstream ``Dipole C'' chamber and a 5.03~m long upstream ``Dipole
A'' chamber, which are integrated with magnet girders, and a 3.69~m
long undulator vacuum chamber.

\begin{figure}
\centering
    \includegraphics[width=0.45\textwidth]{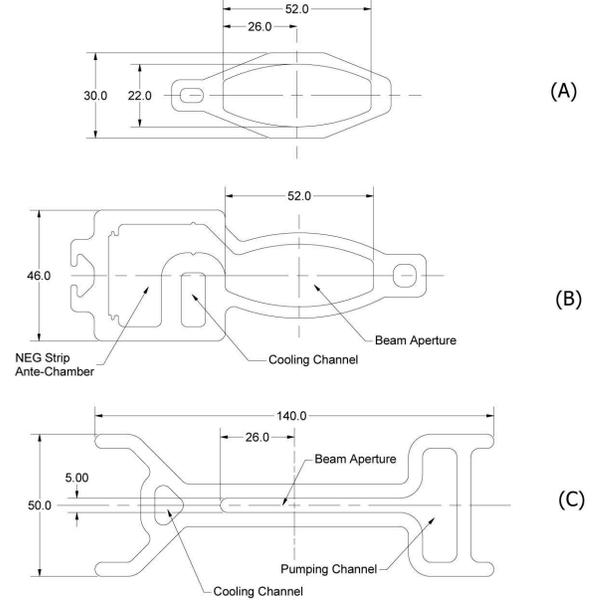}
    \caption{Three types of aluminum extrusions used in construction of CHESS-U vacuum chambers.
(A) Quad extrusions are used in straight sections with tight fit
between magnet poles and beampipes; (B) Dipole extrusions are used
to form bending sections in DQ magnets, which includes ante-chamber
for NEG-strips.  (C) Undulator extrusions are the base material for
the undulator vacuum chambers. }
    \label{fig:vac-extrusion_profiles}
\end{figure}

\begin{figure}[tbh]
    \centering
        \begin{subfigure}[t]{0.45\textwidth}
            \includegraphics[width=\textwidth]{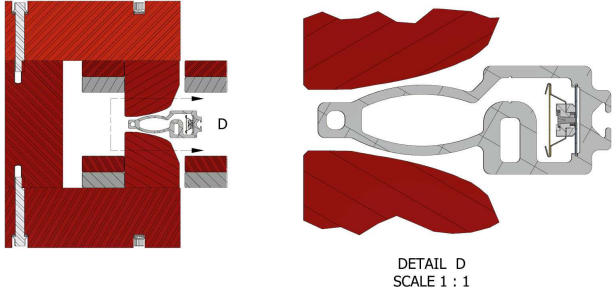}
            \caption{Dipole extrusion in a DQ magnet.}\label{fig:vac-dq}
        \end{subfigure}
        \begin{subfigure}[t]{0.45\textwidth}
            \includegraphics[width=\textwidth]{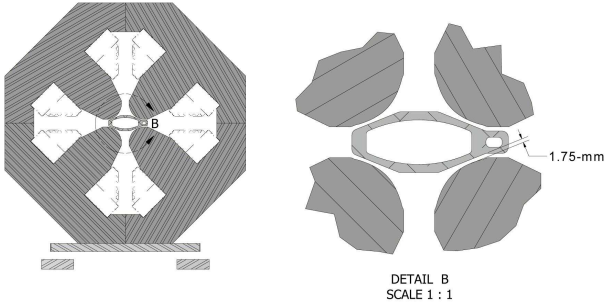}
            \caption{Quadrupole extrusion in a quadrupole magnet.}
            \label{fig:vac-quad}
        \end{subfigure}
        \caption{Cross-sections of extrusions in their respective magnets.}
        \label{fig:vac-cross_section}
    \end{figure}

The majority of the chambers are made from 3 styles of 6063 aluminum
extrusions (Fig. \ref{fig:vac-extrusion_profiles}): dipole
extrusion, quad extrusion and undulator extrusion. The minimum
clearances between magnet pole tips and the extrusions are slightly
less than 2~mm, as illustrated in Fig. \ref{fig:vac-cross_section}.
The dipole and quad extrusions have a beam aperture of 52~mm (H)
$\times$ 22~mm (V), while the vertical aperture in the undulator
extrusion is 5~mm.  The dipole and undulator extrusions have
ante-chambers for distributed pumping.

\begin{figure}
\centering
    \begin{subfigure}[t]{0.45\textwidth}
    \includegraphics[width=\textwidth]{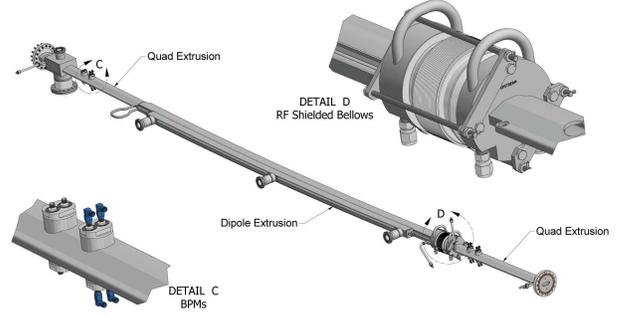}
    \caption{Dipole C vacuum chamber.}\label{fig:vac-dipole_c}
    \vspace*{10mm}
    \end{subfigure}
    \begin{subfigure}[t]{0.45\textwidth}
    \includegraphics[width=\textwidth]{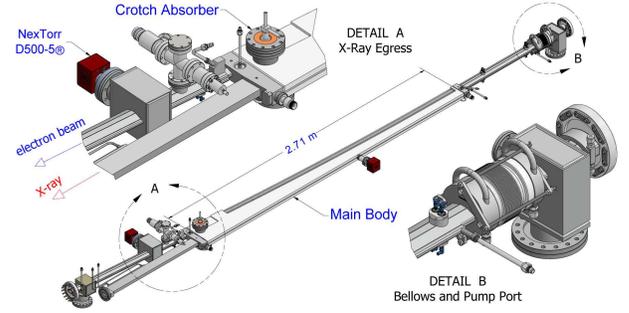}
    \caption{Dipole A vacuum chamber.}\label{fig:vac-dipole_a}
    \vspace*{10mm}
    \end{subfigure}
    \begin{subfigure}[t]{0.45\textwidth}
    \includegraphics[width=\textwidth]{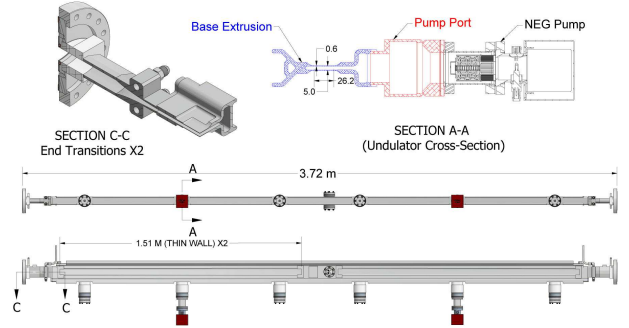}
    \caption{Undulator vacuum chamber.}\label{fig:vac-undulator}
    \vspace*{4mm}
    \end{subfigure}
    \caption{Vacuum chamber detail.}
    \label{fig:vac-detail}
\end{figure}

    The bending section of a dipole C chamber, as shown in Fig. \ref{fig:vac-dipole_c},
    is made from the dipole-style extrusion (Fig. \ref{fig:vac-extrusion_profiles}b).
    Precision bending (31.17~m radius, 4.26 degree bending angle) of the
    extrusion was achieved via the stretch-forming technique, in which
    an aluminum extrusion was formed to a precision die while being
    stretched to its yield stress.  The stretch-formed section was
    then machined to its design form, and the quad-style (Fig. \ref{fig:vac-extrusion_profiles}a)
    extrusions welded to both ends.  Four sets of BPMs were welded directly
    onto the quad extrusions near locations of quadrupole magnets.
     Other functional vacuum components, including an RF-shielded
     sliding joint and pumping ports, were also welded to the quad extrusions.
     The chamber completes with a pair of stainless flanges with
     aluminum-to-stainless steel bi-metal transitions.

    The dipole A chamber design is much more complicated than the dipole C chamber,
    as it needs to provide egress for the X-ray from the canted undulators.
    The dipole A design is illustrated in Fig. \ref{fig:vac-dipole_a}.  To incorporate the compact
    geometry needed for the passage of both stored electron beam and egress of X-ray
    from the undulators, the chamber main body is constructed by welding of two
    precision machined ``clam-shells'' of 6061-T6 aluminum alloy.  Cooling channels
    and NEG strip mounting features are also machined in the main body.  Similar to
    the dipole C chamber, quad extrusions are welded to both ends of the main body.
    Two sets of BPMs, one RF-shielded sliding joint and multiple pump ports are also
    directly welded to the chamber. At separation junction of the electron beam and the
    X-ray, an insertable crotch absorber is mounted onto the dipole A chamber, to
    intercept up to 5.4~kW of SR power from the dipole magnet, while allowing required
    passages of both electron beam and the X-ray.  The crotch absorber is one of the most
    challenging CHESS-U vacuum components, due to very limited available space.
    Its design is based on the existing CHESS crotches, consisting of a pure beryllium
    ring (5~mm thick) and a water-cooled copper core.  The beryllium ring, vacuum brazed
    to the copper core, dilutes the SR power density by bulk absorption and scattering along
    its depth.  FEA analysis indicates that the crotch absorber has sufficient
     margin of safety to handle the SR power from the design CHESS-U beam parameters (6.0~GeV,
     250~mA).
    Design and construction of the CHESS-U undulator chambers are similar to an existing
    CCU chamber in operation for the last four years.  As shown in Fig. \ref{fig:vac-undulator}, the undulator
    beampipe is made from the CHESS-U undulator extrusion (Fig. \ref{fig:vac-extrusion_profiles}c).
    Two sections (1.51~m each) of the extrusion are machined to a thickness of 0.61~mm to accommodate
    the 7.0~mm CCU pole gap.  The machining of the undulator extrusion is done dry to avoid
    difficulty in cleaning the long beampipe with 5~mm vertical aperture with thin walls.
    The wall thickness at the thin section was constantly checked using ultrasonic thickness
     gauges during the machining.  Two end assemblies are welded to the undulator beampipe,
     that gently transition vertically from 5~mm to 22~mm, the nominal CHESS-U vertical aperture.
      Six pumping ports are welded to the ante-chamber of the undulator extrusion.

\subsection{Vacuum Pumping and Performance Simulation}
All CHESS-U vacuum chambers were baked to 150$^\circ$~C and then
back-filled with ultra-high purity nitrogen through a MATHESON
NANOCHEM\textregistered\; purifier that reduces H2O/HC level below
ppb level.  A post-installation 95$^\circ$~C hot water bakeout of
all CHESS-U chambers is also planned.  With these measures,
SR-induced desorption (SRID) is expected to be the dominant source
of gas load.

\begin{figure}
\centering
    \includegraphics[width=0.45\textwidth]{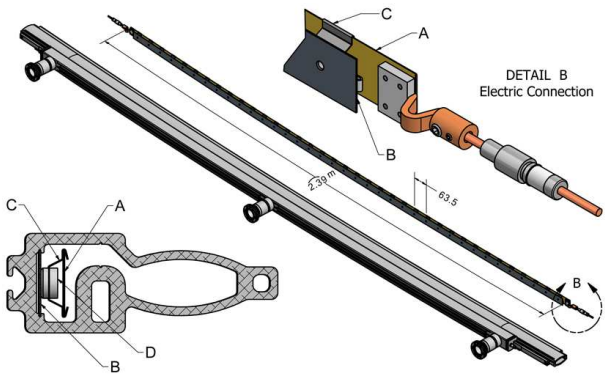}
    \caption{- NEG strip assembly in dipole chambers.
    (A) 30-mm NEG (SAES Getters st-707) strip;
    (B) Stainless steel carrier ribbon; (C) Stainless steel clips; (D) Alumina
    spacers.}
    \label{fig:vac-neg_assembly}
\end{figure}

Owing to distributed nature of the SRID gas load and restricted
conductance of the CHESS-U beampipe, it is essential to have
distributed vacuum pumping.  Non-evaporable getter (NEG) strips
(st-707 SAES Getters) were chosen for the distributed pumps in the
dipole chambers over distributed ion pump in CESR, due to the
limited space available.  The structure of the NEG strip assembly is
shown in Fig. \ref{fig:vac-neg_assembly}.  A 30~mm wide NEG strip is
supported on a stainless steel ribbon with periodic stainless steel
clips.  The clips are pinched to the NEG strip, and are riveted to
the stainless steel ribbon through sets of alumina spacers.  The NEG
strip will be activated by resistive heating to 500$^\circ$~C with a
DC power supply. Flexible connections at both ends of the NEG strip
to the electric vacuum feedthroughs allow thermal expansion during
activation.

Very high SRID gas load is expected in the initial beam conditioning
stage.  Simulations showed that the NEG strips may not have
sufficient pumping capacity during the CHESS-U commissioning phase,
and it can be disruptive to the initial operation with frequent
activation cycles.  Therefore, compact high capacity NEG modular
pumps, CapaciTorr Z200 and NexTorr Z100 (SAES Getters), are
installed to aid the initial beam conditioning.  To handle potential
noble (or non-gettable) gases, a large sputtering ion pump (approx.
110~l/s) is also installed on each achromat cell.

\begin{figure}
\centering
    \includegraphics[width=0.45\textwidth]{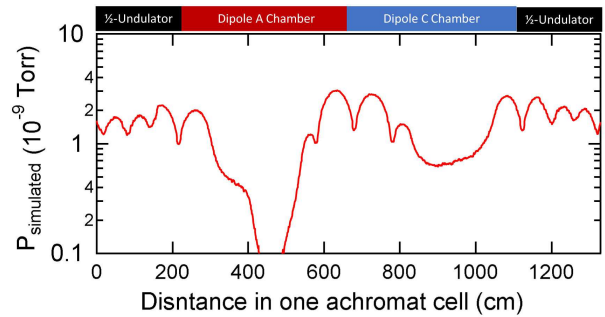}
    \caption{Simulated CHESS-U pressure profiles with 250~mA stored beam
    with 100~A-hr of beam-conditioning, assuming NEG pumps at nominal
    pumping speed, using MolFlow+, which showed an average pressure
    of $\approx 1.8\times 10^{-9}$~torr. }
    \label{fig:vac-molflow}
\end{figure}

The CHESS-U vacuum pumping performance was evaluated using MolFlow+,
a Test-Particle Monte-Carlo simulator developed at CERN. The
simulations showed that the installed vacuum pumping system is
capable of achieving required level of vacuum for CHESS-U
operations. Figure \ref{fig:vac-molflow} displays an example of
simulated pressure profile in one CHESS-U achromat cell, after
100~Amp-hr beam conditioning, achieving an average pressure of
$\approx 1.8\times 10^{-9}$~torr. Pressure bumps are due to lack of
pumping in the straight sections populated by the magnets.

%% file: s6_instrumentation.tex
    CESR is well-instrumented for a multitude of beam-based
    measurement techniques, thanks to the CesrTA
    program. Details are documented elsewhere
    \cite{JINST12:T11006}. A few highlights are noted here.

    \subsection{CESR Beam Position Monitors}
        CESR is presently equipped with 110 peak-detection BPMs of an in-house
        design (CBPM-II), capable of measuring bunch-by-bunch turn-by-turn positions
        for bunch spacings $\ge 4$~ns. 100 CBPMs are used for
        routine optics correction and orbit monitoring. Each
        achromat in CHESS-U will be instrumented with four CBPM-II modules,
        one adjacent to every quadrupole. CBPM-II specifications
        are listed in Table \ref{tab:cbpm-ii}; further details are
        available in \cite{JINST12:T09005}.

        \begin{table}
        \begin{centering}
           \begin{tabular}{cccc}
               \toprule
               \textbf{Parameter}    & \textbf{Specification} \\
               \colrule
               Front-end bandwidth  & 500~MHz\\
               Absolute position accuracy (long-term) & 100~$\mu$m\\
               Single-shot position resolution & 10~$\mu$m \\
               Differential position accuracy & 10~$\mu$m \\
               Channel-to-channel sampling time resolution &
               10~psec\\
               \botrule
           \end{tabular}
           \caption {CBPM-II specifications.}
           \label{tab:cbpm-ii}
       \end{centering}
       \end{table}

        With the installation of canted CCU undulators in 2014
        \cite{doi:10.1063/1.4952782},
        three Libera Brilliance Plus processors have been used for
        continuous turn-averaged position monitoring at the existing
        3.5~m-long 4.6~mm-aperture CCU chamber.

    \subsection{X-Ray Video Beam Position Monitors (vBPMs)}

\begin{figure}
        \centering
        \begin{subfigure}[t]{0.35\textwidth}
            \includegraphics[width=\textwidth]{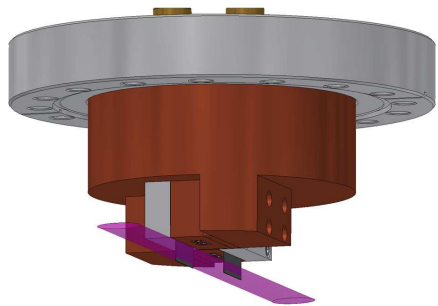}
            \caption{Schematic of diamond blades intercepting
            synchrotron radiation. The blades are separated by
            approximately 7~mm.}
        \end{subfigure}
        \begin{subfigure}[t]{0.35\textwidth}
            \includegraphics[width=\textwidth]{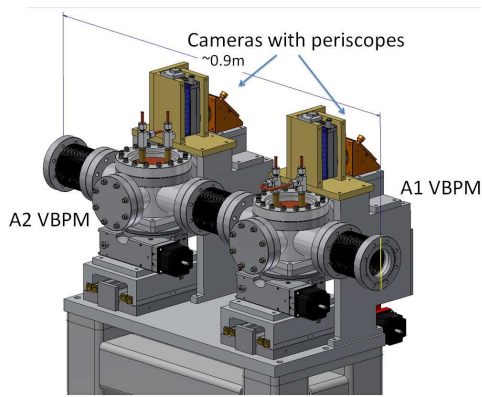}
            \caption{Implementation of two adjacent vBPMs for
            monitoring positions of two canted undulator beams
            separated by 10~mm (1~mrad angular separation).}
        \end{subfigure}
        \caption{vBPM developed for A-line.}
    \label{fig:vbpm}
    \end{figure}

    Each x-ray front end will be equipped with two vBPMs, one
    for each extracted beam, to provide horizontal and vertical
    position. The vBPMs will use an edge-on diamond
    blade monitor which was prototyped on A-line in 2015
    \cite{temnykh:vbpm}, rendered in Fig. \ref{fig:vbpm}.
    The diamond-blade style vBPM allows for a compact design close
    to the source (roughly 10~m), where the canted undulator beams
    are only separated by 10~mm.

        The vBPMs are presently used in conjunction with Libera BPMs
        around the small-gap undulator chamber
        to monitor x-ray trajectories in real-time and
        correct positions after every topoff, typically every five
        minutes. This strategy will be expanded in
        CHESS-U, where each sector will
        have two Libera Brilliance Plus monitors dedicated
        to real-time position monitoring for position feedback.

    \subsection{Turn-by-Turn Feedback}
        Bunch-by-bunch turn-by-turn horizontal, vertical, and longitudinal
        feedback is accomplished using three Dimtel feedback front-end and processing
        modules \cite{DIMTEL2009:IGP1281F:Man17}.
        The low level outputs are connected to 200~W, 150~MHz ENI amplifiers which
        drive horizontal and vertical stripline kickers. The longitudinal processor
        drives a 200~W solid-state RF amplifier that in turn drives a 1~GHz low Q cavity
        kicker.  While the system can handle bunch spacings as short as 2~ns, gain falls
        off because of the short interval required to provide feedback.
        A bunch spacing of 14~ns  provides for the maximum kick as well as
        matching better with the injector RF frequency.

%% file: s7_bunch_patterns.tex
Historically, bunch patterns were constrained by the beam-beam
interaction from electrostatically-separated counter-rotating beams
in the same vacuum chamber. With the simplification of single-beam
operation, there are many more possibilities.

There is increasing interest in time-resolved experiments (see, for
example, \cite{doi:10.1063/1.4893881, naturecomm8:14377}), and a
number of possible bunch patterns are under consideration. The
constraints on bunch patterns are now discussed, and several bunch
patterns proposed.

\subsection{Mode Coupling Instability Threshold}

Many light source upgrades are favoring mult-bend achromat designs
with strong focusing and small vacuum apertures (see, for example,
\cite{APS-U:PDR, ESRF-II:OrangeBook}). However, the resistive wall
impedance scales as $1/r^3$, quickly limiting the single-bunch
current due to mode coupling instability \cite{sachererTMCI}. The
vacuum chamber aperture for CHESS-U has been kept relatively large
(52~mm $\times$ 22.5~mm full-aperture) in an effort to keep the
maximum single bunch current as high as possible.

 The transverse mode coupling
instability (TMCI) threshold has been estimated for CHESS-U based on
analytic resistive wall calculations \cite{PhysRevE.47.656} and
detailed T3P modeling for discontinuities
\cite{ace3p,1742-6596-125-1-012077}. The limit is estimated to be
around 25~mA (64~nC) \cite{CLASSE:REU2017:Salo}; conservatively,
only bunch patterns up to 20~mA are considered.

The present single-bunch current limit is 10~mA (26~nC) per bunch,
to protect instrumentation (BPM modules and feedback amplifiers).
Options for raising the limit are under discussion, opening the path
to high bunch-current operating modes. These options include
preventing damage to the feedback amplifiers by limiting
beam-induced power transmitted back to the amplifier.

\subsection{Momentum Aperture and Touschek Lifetime}

\begin{figure}
\centering
    \begin{subfigure}[t]{0.35\textwidth}
    \includegraphics[width=\textwidth]{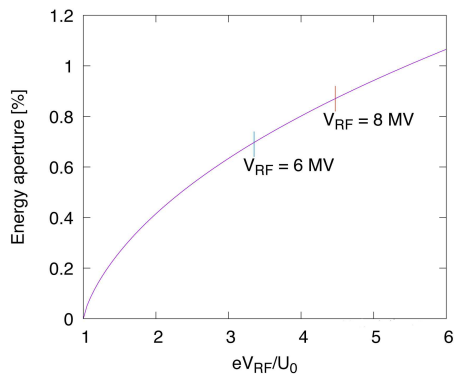}
    \caption{Momentum aperture.}
    \end{subfigure}
    \begin{subfigure}[t]{0.35\textwidth}
    \includegraphics[width=\textwidth]{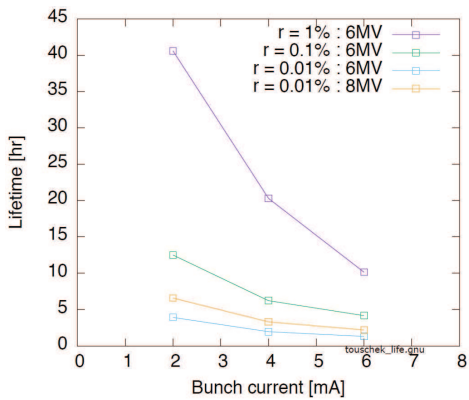}
    \caption{Touschek lifetime.}
    \end{subfigure}
    \caption{Momentum aperture and Touschek lifetime as a function
    of RF voltage, bunch current, and emittance coupling. }
    \label{fig:rf_requirements}
\end{figure}

The momentum compaction factor $\alpha_p$ for CHESS-U is rather
large compared to other modern light sources, resulting in a limited
momentum aperture (Fig. \ref{fig:rf_requirements}, top). The
Touschek lifetime (Fig. \ref{fig:rf_requirements}, bottom) at the
design emittance coupling of 1\%, total RF voltage of 6~MV, and
design bunch current of 2.22~mA (in $18\times 5$ mode) is roughly
40~hrs.

Extrapolating to 25~mA, the anticipated Touschek lifetime at 0.1\%
coupling is approximately one hour; at 1\% coupling, this increases
to around 3 hours.

\subsection{Electron Cloud}

    Initial commissioning will be done with positrons. As such,
    bunch patterns will be limited by tune shift and head-tail
    instability due to the electron cloud effect
    \cite{CornellU2013:PHD:JCalvey}. Each of the bunch patterns
    proposed here were examined for electron cloud build-up
    and trapping \cite{CLASSE:REU2018:KRowan}. The electron cloud
    tune shift was constrained to be less than or equal to
    the electron cloud tune shift in present two-beam bunch
    patterns (around 3.5~kHz, or $\Delta Q \approx 0.009$ in
    both horizontal and vertical). It is possible that larger tune
    shifts from electron cloud would be tolerable.

\subsection{Proposed Bunch Patterns}

Bunch patterns proposed for CHESS-U are listed in Table
\ref{tab:bunch_patterns}. Commissioning and early operation will be
in the $18\times5$ pattern, with the option of exploring other bunch
patterns as user interest dictates.

\begin{table}
 \begin{centering}
     \begin{tabular}{cccc}
         \toprule
         \textbf{Parameter}    &\;\; \textbf{$18\times5$} \;\;   & \;\;\textbf{$9\times5$}\;\;    & \;\;\textbf{$9\times1$}\;\;\\
         \colrule
         \# Trains              & 18                      & 9 & 9 \\
         Train Spacing [ns]    & 84                    & 224  & 280 \\
         \# Bunches / Train     & 5                     & 5  & 1 \\
         Bunch Spacing [ns]    & 14                     & 14  & n/a \\
         Charge/Bunch [mA]     & 2.22                 & 4.44  & 20 \\
         Total current [mA]    & 200                  & 200   & 180 \\
         $\tau_{Touschek}$ at 1\% Coupling [hrs] & 40 & 17 & 4 \\
         \botrule
     \end{tabular}
     \caption {Bunch patterns for CHESS-U. 1~mA = $1.6\times 10^{10}$ particles =
     2.56~nC.}
     \label{tab:bunch_patterns}
 \end{centering}
 \end{table}

In any of the proposed bunch patterns, CESR would run with topoff.
In present CHESS two-beam conditions, the topoff interval for
positrons is 5~min, with a Touschek lifetime of approximately
12~hrs.

CESR will preserve the ability to store electrons in the
counter-clockwise direction for machine studies. Total current in
the counter-clockwise direction will be limited to 5~mA (13~nC) at
6~GeV, limited by synchrotron radiation heating on the sliding
joints in the new achromats. This current limit will scale as
$E^{-4}$; therefore, the restriction in the counter-clockwise
direction at 2.1~GeV (frequently used for machine studies) will
exceed the existing administrative limit of 200~mA.

%% file: s8_commissioning.tex
Installation of the CHESS-U upgrade began in June 2018, and is
scheduled to be completed in November 2018. Beam commissioning will
follow. All insertion devices are out-of-vacuum, which allows for
initial commissioning and vacuum processing without IDs. Undulators
will be installed one canted pair at a time, with a brief recovery
period after each installation to establish operating conditions.
Beam will be delivered to all end stations by April 2019.

The positron beam presently circulates in the clockwise direction;
as such, initial commissioning and operation of CHESS-U will be with
positrons, with the option to invert the polarity of the injector
and storage ring to run electrons in the clockwise direction at a
future date.

%% file: appendices.tex

 \label{sec:error_table}\section{Misalignments for Error
Tolerance}

Standard deviations for misalignments applied in error tolerance
simulations are listed in Table \ref{tab:ringma_errors}. The errors
for dipoles, quadrupoles, and sextupoles are based on survey
information. DQ and girder errors are determined by extrapolation
from survey information and magnetic alignment characterization.

BPM misalignments and errors are listed in Table
\ref{tab:bpm_errors}.

\begin{table}
 \begin{centering}
      \caption {RMS error amplitudes for misalignment and correction simulation.}
     \label{tab:ringma_errors}
     \begin{tabular}{cccccc}
         \toprule \rule{0pt}{2.5ex}
         \textbf{Error}    & \textbf{Dipole}    & \textbf{DQ} & \textbf{Quad} & \textbf{Sext} & \textbf{Girder}\\
         \colrule \rule{0pt}{2.5ex}
         x-offset [$\mu$m]         & 2000       &     100  &   200  &   300  & 100 \\
         y-offset [$\mu$m]         & 900        &     100  &   110  &   300  & 100 \\
         s-offset [mm]             & 2.3        &     2.0  &   5.2  &   5.2  & 0.1 \\
         Pitch [$\mu$rad]          & 600        &     87   &   1100 &   1200 & 100 \\
         Yaw   [$\mu$rad]          & 300        &     87   &   62   &   800  & 100 \\
         Roll  [$\mu$rad]          & 500        &     500  &   500  &   500  & 100 \\
         Gradient [\%]             &  --        &     0.1  &   0.1  &   0.1  &  -- \\
         \botrule
     \end{tabular}
 \end{centering}
 \end{table}

 \begin{table}
 \begin{centering}
      \caption {RMS BPM errors for misalignment and correction simulation.}
     \label{tab:bpm_errors}
     \begin{tabular}{cc}
         \toprule \rule{0pt}{2.5ex}
         \textbf{Error} & \textbf{Amplitude} \\
         Transverse offset [$\mu$m]& 170 \\
         Reproducibility [$\mu$m]  & 10  \\
         Roll [mrad]               & 12  \\
         Timing  [ps]              & 10  \\
         Shear [$\mu$m]            & 100 \\
         Channel Gain [\%]         & 0.5 \\
         \botrule
     \end{tabular}
 \end{centering}
 \end{table}

\label{sec:multipoles} \section{Systematic Multipoles}

The multipole expansion is parameterized as a fractional field error
$dB/B$ with normal ($b_n$) and skew ($a_n$) components, where $n=1$
corresponds to a quadrupole-like field. The reference radius for all
multipoles listed in Table \ref{tab:systematic_multipoles} is 20~mm.
All multipoles are normalized to the main field.

    \begin{table}
    \begin{centering}
        \caption{Systematic multipoles used in tracking simulations.
        ``CESR'' magnets are those which are outside the new CHESS-U
        achromats. Evaluated as $dB/B$ at a reference radius of 20~mm.}
        \label{tab:systematic_multipoles}
        \begin{tabular}{ccr}
            \toprule \rule{0pt}{2.5ex}
            \textbf{Element Class} &
            \textbf{Multipole} &
            \textbf{Amplitude} \\
            \colrule \rule{0pt}{2.5ex}
            CESR Dipole         & $b_1$    & $-0.5\times 10^{-4}$ \\
                                & $b_2$    & $-0.2\times 10^{-4}$ \\
                                & $b_{12}$ & $-2.4\times 10^{-9}$ \\
            \colrule \rule{0pt}{2.5ex}
            CESR Quadrupole     & $b_5$    & $1.5\times 10^{-4}$ \\
                                & $b_9$    & $-0.053\times 10^{-4}$ \\
            \colrule \rule{0pt}{2.5ex}
            CESR Sextupole      & $b_8$    & $1\times 10^{-4}$ \\
            \colrule \rule{0pt}{2.5ex}
            CHESS-U Quads A/D   & $b_2$ & $ 1 \pm 6.5 \times 10^{-4}$\\
                                & $b_3$ & $ 0.4 \pm 1.7 \times 10^{-4}$\\
                                & $b_4$ & $ -0.3 \pm 1.5 \times 10^{-4}$\\
                                & $b_5$ & $ -4 \pm 1.5 \times 10^{-4}$\\
                                & $a_2$ & $ 0.8 \pm 7.8 \times 10^{-4}$\\
                                & $a_3$ & $ -0.3 \pm 2.8 \times 10^{-4}$\\
                                & $a_4$ & $ -0.5 \pm 1.9 \times 10^{-4}$\\
                                & $a_5$ & $ 0.1 \pm 0.6 \times 10^{-4}$\\
            \colrule \rule{0pt}{2.5ex}
            CHESS-U Quads B/C   & $b_2$ & $ 1.2 \pm 6.1 \times 10^{-4}$\\
                                & $b_3$ & $ 0.8 \pm 2.3 \times 10^{-4}$\\
                                & $b_4$ & $ 0.8 \pm 1.8 \times 10^{-4}$\\
                                & $b_5$ & $ -2.6 \pm 2.1 \times 10^{-4}$\\
                                & $a_2$ & $ -1.7 \pm 3.3 \times 10^{-4}$\\
                                & $a_3$ & $ -2.6 \pm 4.0 \times 10^{-4}$\\
                                & $a_4$ & $ -2.2 \pm 2.9 \times 10^{-4}$\\
                                & $a_5$ & $ 0.3 \pm 0.7 \times 10^{-4}$\\
            \colrule \rule{0pt}{2.5ex}
            CHESS-U DQ          & $b_2$ & $-3.4 \pm 5.6 \times 10^{-4}$ \\
                                & $b_3$ & $-2.3 \pm 1.9 \times 10^{-4}$ \\
                                & $a_1$ & $-1 \pm 2.8 \times 10^{-4}$ \\
                                & $a_2$ & $-0.6 \pm 2.6 \times 10^{-4}$ \\
                                & $a_3$ & $0.1  \pm 1.4 \times 10^{-4}$ \\
            \botrule
        \end{tabular}
    \end{centering}
    \end{table}